\documentclass[twocolumn,twocolappendix]{aastex631}
\usepackage{apjfonts}
\usepackage{CJK}

\usepackage{amsmath}
\usepackage{comment}

\usepackage{graphicx}
\usepackage[utf8]{inputenc}

\newcommand{\Tfrac}[2]{\left(\frac{#1}{#2}\right)}

\DeclareMathAlphabet\mathbfcal{OMS}{cmsy}{b}{n}

\newcommand{\hatin}[1]{_{\hat{#1}}}

\usepackage{lipsum}

\newcommand{\mean}[1]{\left\langle #1 \right\rangle}
\newcommand{\phimean}[1]{\left\langle #1 \right\rangle_\varphi}
\newcommand{\thmean}[1]{\left\langle #1 \right\rangle_\theta}

\begin{document}

\title{Demystifying flux eruptions:\\ Magnetic flux transport in magnetically arrested disks}

\author[0000-0003-2982-0005]{Jonatan Jacquemin-Ide}
\email{Jonatan.Jacquemin@colorado.edu}
\affiliation{JILA, University of Colorado and National Institute of Standards and Technology, 440 UCB, Boulder, CO 80309-0440, USA}

\author[0000-0003-0936-8488]{Mitchell C. Begelman}
\affiliation{JILA, University of Colorado and National Institute of Standards and Technology, 440 UCB, Boulder, CO 80309-0440, USA}
\affiliation{Department of Astrophysical and Planetary Sciences, University of Colorado, 391 UCB, Boulder, CO 80309-0391, USA}

\author[0000-0002-2875-4934]{Beverly Lowell}
\affiliation{Center for Interdisciplinary Exploration $\&$ Research in Astrophysics (CIERA), Physics and Astronomy, Northwestern University, Evanston, IL 60202, USA}

\author[0000-0003-4475-9345]{Matthew Liska}
\affiliation{Center for Relativistic Astrophysics, Georgia Institute of Technology, Howey Physics Bldg, 837 State St NW, Atlanta, GA, 30332, USA}

\author[0000-0003-3903-0373]{Jason Dexter}
\affiliation{JILA, University of Colorado and National Institute of Standards and Technology, 440 UCB, Boulder, CO 80309-0440, USA}
\affiliation{Department of Astrophysical and Planetary Sciences, University of Colorado, 391 UCB, Boulder, CO 80309-0391, USA}

\author[0000-0002-9182-2047]{Alexander Tchekhovskoy}
\affiliation{Center for Interdisciplinary Exploration $\&$ Research in Astrophysics (CIERA), Physics and Astronomy, Northwestern University, Evanston, IL 60202, USA}

\begin{abstract}
Magnetically arrested disks (MADs) are a compelling model for explaining variability in low-luminosity active galactic nuclei, including horizon-scale outbursts like those observed in Sagittarius A*. MADs experience powerful flux eruptions—episodic ejections of magnetic flux from the black hole horizon—that may drive the observed luminosity variations. In this work, we develop and validate a new formalism describing large-scale magnetic field transport in general relativistic magnetohydrodynamic simulations of MADs with geometrical thicknesses of $h/R=0.1$ and $h/R=0.3$. We introduce a net flux transport velocity, $v_\Phi$, which accounts for both advective and diffusive processes. We show that MADs maintain a statistical quasi-steady state where advection and diffusion nearly balance. Flux eruptions appear as small deviations from this equilibrium, with $v_\Phi/V_k  \ll 1$, where $V_k$ is the local Keplerian velocity. Using this framework, we analytically derive a recurrence timescale for flux eruptions, $t_{\rm rec} \simeq 1500\,\,r_g/c$. This timescale closely matches simulation results. The smallness of $v_\Phi$ explains the long recurrence times of flux eruptions compared to other system timescales. 

We also take a closer look at the diffusion of the magnetic field by performing the first measurement of turbulent resistivity in MADs. We then estimate the turbulent magnetic Prandtl number, $\mathcal{P}_m$, defined as the ratio of turbulent viscosity to turbulent resistivity. We find $\mathcal{P}_m \simeq 1$–$5$, consistent with shearing-box simulations of magneto rotational instability-driven turbulence. While flux eruptions excite large-scale non-axisymmetric modes and locally enhance turbulent resistivity, magnetic field diffusion is dominated by smaller-scale turbulent motions. These results provide new insight into the nature of AGN variability and the fundamental physics of magnetic field transport.
\end{abstract}

\keywords{ accretion, accretion disks --- magnetohydrodynamics (MHD) ---  stars: black holes --- galaxies: active}


\section{Introduction} \label{sec:intro}


Despite extensive study, many aspects of accreting systems, including their variability and episodic outbursts, remain poorly understood. These variable events often occur on timescales comparable to the inner disk’s dynamical timescales, providing rare insights into the physics of these extreme environments. Understanding their origin is critical, as variability offers one of the few signatures of the inner workings of accreting compact objects. Sagittarius A* is known to exhibit near-infrared flares \citep{genzel_near-infrared_2003}, accompanied by X-ray flares \citep{baganoff_rapid_2001,marrone_x-ray_2008}, yet their origin remains debated. These flares originate from compact regions near the BH, rotating at roughly Keplerian frequencies \citep{abuter_detection_2018}. Many AGN, including M87, also produce gamma-ray flares, which likely result from particle acceleration processes, though the precise mechanism remains unknown \citep{aharonian_fast_2006,albert_variable_2007,aliu_veritas_2012,aleksic_black_2014}. These gamma-ray flares are sometimes accompanied by X-ray emission, which would link these flares to a compact emission sources near the central BH \citep{sun_energy_2018}. Therefore, we focus on gamma-ray flares originating near the event horizon, likely associated with magnetic reconnection within the black hole magnetosphere, rather than those produced on larger jet scales.

A correlation between observed jet power and accretion power has been established in X-ray binaries and AGN, spanning multiple orders of magnitude in luminosity \citep{corbel_radio/x-ray_2003,corbel_universal_2013,zamaninasab_dynamically_2014}. This correlation highlights the close link between accretion and ejection as interdependent processes. Dynamically, both are tied to large-scale magnetic fields, which play a central role in shaping jet launching and accretion. Jets require large-scale vertical magnetic flux anchored near the black hole’s event horizon \citep{blandford_electromagnetic_1977}. Accretion, in turn, is facilitated by turbulence—driven by the magnetorotational instability \citep[MRI, ][]{balbus_powerful_1991,balbus_instability_1998} in the presence of a vertical magnetic field\footnote{Though the initial field could also be small-scale or toroidal \citep{hawley_local_1995}}—or by large-scale torques (or wind torques) enabled by the large scale vertical field through the \cite{blandford_hydromagnetic_1982} mechanism, i.e. MHD disk winds extract angular momentum from the disk \citep{ferreira_magnetized_1995}.

However, how magnetic flux is transported to the inner regions of the accretion disk, near the central BH engine, remains largely unknown. Furthermore, due to the no-hair theorem, black holes tend to expel magnetic fields, meaning the field must be constantly replenished. Two scenarios are typically considered: magnetic field advection from larger scales or in-situ field generation via a dynamo process within the disk.
The latter scenario will not be explored in this work, though it has been observed in global simulations \citep{liska_large-scale_2020,jacquemin-ide_magnetorotational_2024,kaaz_h-amr_2025}. However, even when a dynamo operates, advection of magnetic flux from larger radii into the inner disk appears to play a crucial role in maintaining the magnetic field structure \citep{jacquemin-ide_magnetorotational_2024,kaaz_h-amr_2025}.

Whether the advection of a large-scale magnetic field is possible has long been an open question. This issue is crucial, as not all accreting systems produce jets—only about $10\%$ of AGN are jetted \citep{padovani_microjansky_2011,zamaninasab_dynamically_2014}, XRBs launch jets only during specific outburst phases \citep{done_modelling_2007}, and only a few TDEs generate prompt jets \citep{burrows_relativistic_2011}. Thus, understanding the conditions required for jet formation necessarily involves understanding how magnetic fields are transported within accretion flows.

In the context of viscous accretion disks, the seminal work of \cite{lubow_magnetic_1994} demonstrated that the ability of a magnetic field to be transported inward depends critically on the properties of the accretion disk. They showed that inward advection occurs only when $D=(R/h)(\eta/\nu)\lesssim1$, where $R$ is the disk radius, $h$ is the disk’s geometrical thickness, and $\eta$ and $\nu$ are the effective turbulent diffusivities for the magnetic field and angular momentum, respectively. They concluded that field advection would be impossible in thin disks if turbulent viscosity is far larger than turbulent resistivity. Later studies of MRI turbulence confirmed that $\eta/\nu \simeq 3$, making the likelihood of field transport in the \cite{lubow_magnetic_1994} framework even more unlikely \citep{guan_turbulent_2009,lesur_turbulent_2009,fromang_turbulent_2009}. The work of \cite{lubow_magnetic_1994} has been criticized for its simplified treatment of turbulent transport, and for neglecting large-scale effects such as magnetized wind torques \citep{ferreira_magnetized_1995,scepi_magnetic_2020}.

\cite{rothstein_advection_2008} pointed out that vertical structure, largely neglected in earlier analyses, could aid magnetic field advection by creating a small, non-turbulent boundary layer where advection is more efficient. They also noted that an additional source of angular momentum transport, such as large-scale \cite{blandford_hydromagnetic_1982}-like torques—neglected by \cite{lubow_magnetic_1994}—could increase the radial advection velocity beyond that driven by turbulence alone, making magnetic field advection more feasible.
\cite{guilet_transport_2012,guilet_transport_2013} demonstrated that even in a purely viscous disk, magnetic field advection is possible; however, they also found that for very thin disks, $h/R\ll1$, and highly magnetized disks, achieving inward advection is hard or impossible within their framework. They point out that some kind of coronal accretion not dissimilar to the one proposed by \cite{rothstein_advection_2008} and observed in simulations by \cite{beckwith_transport_2009} could help with this problem. Subsequent analytical work has show this to be an important effect \citep{li_large-scale_2021}.  \cite{begelman_simple_2024} developed a simplified slab model in which angular momentum is primarily removed through the disk’s surface layers, with the vertical structure playing a key role in the large-scale transport of magnetic flux. Within this model, they found that under certain conditions, runaway advection of large-scale magnetic flux can occur, leading to a final state similar to that seen in global simulations of magnetized accretion disks \citep{tchekhovskoy_efficient_2011}. However, they also find avection to be impossible for razor thin disks, $h/R\ll0.1$.

The first global general relativistic magnetohydrodynamic (GRMHD) simulations of accreting black holes demonstrated that magnetic field advection was not only possible, but a natural outcome of the system’s evolution \citep{beckwith_transport_2009}. In these simulations, the magnetic flux was efficiently advected inward, eventually saturating at its maximal value on the BH horizon \citep{tchekhovskoy_efficient_2011,mckinney_general_2012}. This saturation led to a magnetically arrested state (MAD), in which the excess magnetic flux was ejected outward via powerful flux eruptions from the BH's magnetosphere. 

Weakly magnetized accretion disks also show clear signs of magnetic field advection. Using numerical MHD simulations of accretion disks \cite{jacquemin-ide_magnetic_2021} found that all their initially weakly magnetized disks advected magnetic flux. They measured the rate at which weakly magnetized disks advect magnetic flux and found that stronger initial magnetic fields lead to faster advection. Furthermore, thinner disks exhibit slower advection, which becomes almost undetectable for $h/R=0.05$, consistent with the analytical results of \cite{lubow_magnetic_1994} and the numerical findings of \cite{mishra_strongly_2020}, who observed no advection in similarly thin disks.
For sufficiently strong initial magnetic fields, the disks transition into a highly magnetized state resembling a MAD, albeit in simulations without GR, with flux eruptions appearing from the inner regions \citep{jacquemin-ide_magnetic_2021}. 

The numerically discovered flux eruptions have been a useful framework for understanding the variability in many accretion disk transients. Indeed, flux eruptions propagating through the disk could be loaded with nonthermal positron-electron pairs, potentially powering X-ray transient emission \citep{dexter_sgr_2020,porth_flares_2021,scepi_sgr_2022,zhdankin_particle_2023}. This has important observational implications, as shown by the GRAVITY collaboration, which reports the outburst generating structures rotating at quasi-Keplerian speeds and featuring strong vertical magnetic fields \citep{jimenez-rosales_dynamically_2020}. Showing, a connection between large scale MHD accretion physics and the flares of Sgr A*. This is consistent with EHT observations of M87, which reveal a highly magnetized accretion disk threaded by a large-scale vertical magnetic field capable of launching jets and supporting flux eruptions, resembling a MAD \citep{collaboration_first_2021}.

Due to their observational importance, much work has focused on understanding the dynamics of flux eruptions. They are now widely accepted to be associated with magnetic reconnection within the black hole magnetosphere \citep{ripperda_black_2022}. They are promising sites of particle acceleration and could explain the gamma-ray flares observed in M87 and other AGN  \citep{hakobyan_radiative_2023}. Furthermore, they may drive Rayleigh–Taylor or interchange instabilities \citep{spruit_interchange_1995} due to the large density contrasts surrounding them, which could in turn lead to additional particle acceleration \citep{zhdankin_particle_2023}. Finally, some studies suggest that flux eruptions may also play a role in the transport of angular momentum, although the exact mechanism or instability responsible remains uncertain \citep{chatterjee_flux_2022,most_magnetically_2024}.

Yet, the large-scale dynamics of flux eruptions and their connection to magnetic field transport remain poorly understood. In particular, the origin of their recurrence timescales, typically $t_{\rm rec}\sim1000\,\,r_g/c$, is unknown—despite this being one of their main agreements with horizon scale observations of Sagittarius A* \citep{dexter_sgr_2020}. In this work, our objective is to employ high-resolution GRMHD simulations of MADs to better understand flux transport and its interplay with flux eruptions. To achieve this, we also take a detailed look at accretion disk turbulence—most notably, by computing for the first time the turbulent resistivity in MADs.

The manuscript is structured as follows: in Section~\ref{sec:num_methods}, we introduce the code and simulations used within this work; in Section~\ref{sec:Flux_trans_frame}, we present and validate the analysis framework employed to study flux transport. In Section~\ref{sec:flux_trans_results}, we analyze the phenomenology of flux transport and examine its radial and vertical structure, which allows us to formulate an analytical theory that accurately reproduces the recurrence timescale of flux eruptions. In Section~\ref{sec:turbulent_struc}, we study the turbulent structure responsible for driving the diffusion of the large-scale magnetic field. Finally, in Section~\ref{sec:conc_disc}, we conclude and discuss the theoretical and observational implications of this work.

\section{Methods}
\label{sec:num_methods}

We perform our simulations using the \texttt{H-AMR} code \citep{Liska2022}, which solves the equations of ideal general relativistic magnetohydrodynamics (GRMHD) on a spherical polar grid ($r$, $\theta$, $\varphi$) in Kerr-Schild coordinates. \texttt{H-AMR} is a 3D, GPU-accelerated GRRMHD code built upon the 3D \texttt{HAMRPI} \citep{ressler_electron_2015,ressler_disc-jet_2017} and the earlier 2D \texttt{HARM2D} code \citep{gammie_harm_2003, noble_primitive_2006}. For convenience, we also define the cylindrical radius $R$ and the vertical height $z$.

All simulations use dimensionless units where $G = M = c = 1$, with $M$ being the black hole mass. In these units, the gravitational radius becomes $r_{\rm g} = GM/c^2 = 1$. Magnetic fields are expressed in Lorentz-Heaviside units, so that the magnetic pressure is given by $b^2/2$, where $b$ is the magnetic field measured in the fluid frame.
To ensure that the black hole exterior is causally disconnected from the inner radial boundary, we place at least five radial cells inside the event horizon. Boundary conditions are set as follows: outflow conditions at both radial boundaries, transmissive conditions at the polar boundaries, and periodic conditions in the $\varphi$-direction \citep{liska_h-amr_2022}.

In this work, we reanalyze two simulations of magnetically arrested disks (MADs). The first is a thick, adiabatic MAD around a nearly maximally spinning black hole ($a = 0.9$), originally presented by \cite{liska_large-scale_2020}. The second is a thin, cooled MAD with $h/R = 0.1$, around a more moderately spinning black hole ($a = 0.3$), first introduced in \cite{lowell_evidence_2025}.

The thick disk is initialized as a sub-Keplerian torus following \cite{chakrabarti_1985}, with specific angular momentum $l \propto R^{1/4}$. It has an inner edge at $r_{\rm in} = 6\,r_g$, a pressure maximum at $r_{\rm max} = 13.792\,r_g$, and an outer edge at $r_{\rm out} \approx 4 \times 10^4\,r_g$. The grid extends to $r = 10^5\,r_g$, and the inner boundary lies inside the event horizon. We use a polytropic equation of state with $\gamma = 5/3$, yielding $h/R \sim 0.2$ at $r_{\rm max}$ and $\sim 0.5$ at $r_{\rm out}$.
The thick disk simulation uses a base resolution of $1872 \times 624 \times 128$, uniform in $\log r$, $\theta$, and $\varphi$. Three levels of static mesh refinement in $\varphi$ increase the effective azimuthal resolution to $1024$ near the equator, giving $70\text{--}90$ cells per scale height. The MRI is seeded with a uniform toroidal magnetic field with initial plasma beta $\beta_{\rm ini} = 5$. Although initialized this way, the simulation evolves into a MAD state \citep{manikantan_winds_2023}. See more details of the setup and angular momentum transport of this MAD in \cite{manikantan_winds_2023}.

For the thin disk with $h/R = 0.1$, we apply a cooling source term following \citep{noble_direct_2009} to maintain the target scale height on a local Keplerian timescale. The initial conditions consist of an equilibrium hydrodynamic torus \citep{fishbone_relativistic_1976} with an inner radius of $r_\mathrm{in} = 20\,r_\mathrm{g}$ and a pressure maximum at $r_\mathrm{max} \sim 41\, r_\mathrm{g}$. We slightly adjust $r_\mathrm{max}$ to ensure that the torus extends to a very large, but finite, outer radius of $r_\mathrm{out} \sim  10^4\,r_\mathrm{g}$. The magnetic field is initialized such that the minimum gas-to-magnetic pressure ratio is $\min\beta = 100$. Further details on the initial setup can be found in \cite{lowell_evidence_2025}.
In this work, we reanalyze the high-resolution simulation, named \texttt{H1a0.3hr}, from \cite{lowell_evidence_2025}, which uses a base grid of $512 \times 288 \times 256$ with one level of static mesh refinement (SMR) applied in the region $|\theta - \pi/2| \leq 0.315$ and $4\,r_\mathrm{g} \leq r \leq 500\,r_\mathrm{g}$. This results in an effective resolution of $1024 \times 576 \times 512$ within the disk body.

\section{Flux transport framework}
 \label{sec:Flux_trans_frame}
\subsection{Theory of flux transport}
 \label{sec:theory}
Inspired by the MHD work of \citet{guilet_transport_2012}, we derive a formalism for magnetic field transport in the GRMHD limit. Since our formulation of the flux advection equation differs from theirs—and such a derivation has not yet been presented in the GRMHD context—we provide the full derivation here.

We begin by defining the electromagnetic tensor,
\begin{equation}
    \mathcal{F}_{\mu\nu} = \partial_\mu A_\nu - \partial_\nu A_\mu,
\end{equation}
where $A_{\mu}$ is the four-vector potential. The dual of the electromagnetic tensor is given by
\begin{align}
    {}^{*}\mathcal{F}_{\mu\nu} &= \frac{1}{2} \epsilon_{\mu\nu\kappa\lambda} \mathcal{F}^{\kappa\lambda}, \\
    {}^{*}\mathcal{F}^{\mu\nu} &= - \frac{1}{2} \epsilon^{\mu\nu\kappa\lambda} \mathcal{F}_{\kappa\lambda},
\end{align}
where the second line follows from our convention for the Levi-Civita fully antisymmetric tensor \citep[see][]{gammie_harm_2003}. The magnetic three-vector is then defined as $B^i = {}^{*}F^{it}$, where Latin indices run from 1 to 3. 

We now rewrite the induction equation as a magnetic field transport equation. Starting from Maxwell's equations
\begin{equation}
    \nabla_{\nu} {}^{*}\mathcal{F}^{\mu\nu} = 0,
\end{equation}
we recover the induction equation in conservative form:
\begin{equation}
    \partial_t B^i + \frac{1}{\sqrt{-g}} \partial_j \left[ \sqrt{-g}(b^j u^i - b^i u^j) \right] = 0.
\end{equation}
This can also be expressed as
\begin{equation}
\label{eq:def_indc}
    \partial_t B^i - \epsilon^{tijk} \partial_j \left( \epsilon_{tklm} u^l b^m \right) = 0.
\end{equation}

Next, we relate the magnetic field to the vector potential. Using the definition
\begin{equation}
\label{eq:Bi_defA}
    B^i = {}^{*}F^{it} = \epsilon^{tijk} \partial_j A_k = -\frac{1}{\sqrt{-g}} \tilde{\epsilon}^{tijk} \partial_j A_k,
\end{equation}
where $\tilde{\epsilon}^{tijk}$ is the fully antisymmetric Levi-Civita symbol, we substitute into Eq.~\eqref{eq:def_indc} to obtain
\begin{equation}
    \epsilon^{tijk} \partial_j \left[ \partial_t A_k - \epsilon_{tklm} u^l b^m \right] = 0.
\end{equation}
This implies that the quantity $\partial_t A_k - \epsilon_{tklm} u^l b^m$ is curl-free. By the Helmholtz decomposition theorem, we can write
\begin{equation}
\label{eq:vector_pot}
    \partial_t A_k - \epsilon_{tklm} u^l b^m = \partial_k \mathcal{G},
\end{equation}
where $\mathcal{G}$ arises to satisfy the gauge condition.

To separate this equations into turbulent and average contributions, we apply Reynolds averaging by decomposing a generic quantity $X$ as
\begin{equation}
    X = \phimean{X} + \delta X,
\end{equation}
where $\phimean{X}$ is the mean (large-scale) component and $\delta X$ is the fluctuating (turbulent) part. By construction, $\phimean{\delta X} = 0$, though nonlinear correlations like $\phimean{\delta X \delta Y}$ are generally non-zero and represent turbulent feedback on the mean field. In what follows, $\phimean{}$ denotes an azimuthal ($\varphi$) average.

Applying this Reynolds average to the $\varphi$-component of Eq.~\eqref{eq:vector_pot} will yield a term of the form $\phimean{\partial_\varphi \mathcal{G}}$. While it's trivial to show that $\phimean{\partial_\varphi X} = 0$ for physical quantities $X$, this is not necessarily true for gauge functions due to their non-uniqueness\footnote{We thank Geoffroy Lesur for pointing this out.}. Luckily, as we will show in Section~\ref{sec:validation}, $\phimean{\partial_\varphi \mathcal{G}} \simeq 0$ is a an accurate approximation and thus we assume $\phimean{\partial_\varphi \mathcal{G}} = 0$ within this work.

Unencumbered by the gauge we proceed by Reynolds averaging the $\varphi$-component of Eq.~\eqref{eq:vector_pot}
\begin{equation}
    \partial_t \phimean{A_\varphi} + \sqrt{-g}{E^\varphi} = 0,
\end{equation}
where
\begin{equation}
\label{eq:indc_inter}
    {E^\varphi} = \mathcal{A}^\varphi+\mathcal{E}^\varphi,
\end{equation}
and
\begin{align}
    \mathcal{E}^\varphi &= \phimean{\delta u^r \delta B^\theta} - \phimean{\delta u^\theta \delta B^r}\,,\\
    \mathcal{A}^\varphi &= \phimean{u^r} \phimean{B^\theta} - \phimean{u^\theta} \phimean{B^r}\,,
\end{align}
$\mathcal{E}^\varphi$ is the turbulent electromotive force (EMF), capturing the effect of turbulent fluctuations on the large-scale magnetic field. The term $\mathcal{A}^\varphi$ models the mean (or large scale) effects of transport unto the large scale magnetic field. 

We use the $\varphi$-average of the $\theta$-component of Eq.~\eqref{eq:Bi_defA} to find
$\phimean{B^{\theta}} = \frac{1}{\sqrt{-g}}\partial_r\phimean{A_\varphi}$.
Combining this with Eq.\ref{eq:indc_inter} yields
\begin{equation}
\label{eq:adv_vert_loc}
\partial_t \phimean{A_\varphi} + \frac{1}{\phimean{B^{\theta}}}\left( \mathcal{A}^\varphi+\mathcal{E}^\varphi\right)\partial_r\phimean{A_\varphi} = 0,
\end{equation}
which is the vertically local flux transport equation, where $\mathcal{A}^\varphi$ advects the magnetic field inward and $\mathcal{E}^\varphi$ diffuses it outward. In practice, we will use this equation only when examining the vertical profiles of $\mathcal{A}^\varphi/\phimean{B^{\theta}}$ and $\mathcal{E}^\varphi/\phimean{B^{\theta}}$.

If we want to examine the radial structure, we need a $\theta$-averaged version of Eq.~\eqref{eq:adv_vert_loc}. While this might seem trivial, it involves subtle complications: as shown by previous analytical and semi-analytical work, the transport of large-scale fields depends on the complete vertical structure of the accretion disk \citep{lubow_magnetic_1994,guilet_transport_2013,begelman_simple_2024}. Fortunately, as we show below, the largest contributions of the advection and diffusion terms are concentrated near the disk midplane (see Section~\ref{sec:radvert_struc}). Therefore, we can simply $\theta$-average the different terms within the disk and use them as source terms for an evolution equation of the $\theta$-integrated large-scale flux.

One can follow a similar procedure to the one detailed above to find the $\theta$-integrated version of Eq.~(\ref{eq:indc_inter}):
\begin{equation}
\label{eq:thm_adveq}
\partial_t{\Phi} + v_\Phi\partial_r{\Phi} = 0,
\end{equation}
where
\begin{equation}
\Phi(r,t) = 2\pi A_\varphi(r,\theta=\pi/2) = 2\pi \int\limits_0^{\pi/2}\phimean{ B^{r}} \sqrt{-g}\,\mathrm{d} \theta,
\label{Eq:pol_flux}
\end{equation}
is the magnetic flux passing through half-spheres, and the advection velocity is given by
\begin{equation}
\label{eq:vphi_defs}
v_\Phi = \frac{1}{\thmean{B^\theta}}\thmean{\mathcal{E^\varphi}+\mathcal{A^\varphi}},
\end{equation}
with the latitudinal average defined as
\begin{equation}
\thmean{X} = \frac{1}{\int_{\theta_1}^{\theta_2}\sqrt{-g}\,\mathrm{d}\theta} \int_{\theta_1}^{\theta_2}\sqrt{-g}\,\phimean{X}\,\mathrm{d}\theta,
\end{equation}
where $\theta_1$ and $\theta_2$ are chosen to be a few scale heights above and below the disk.

Note that an approximation for $\Phi$ was used in deriving Eq.~\eqref{eq:thm_adveq}; more details are given in Appendix~\ref{A:deriv_thmean}. However, as shown in Section~\ref{sec:validation}, this approximation does not lead to noticeable deviations in the accuracy of Eq.~\eqref{eq:thm_adveq}, which remains well validated.

Equation~\eqref{eq:thm_adveq} might appear to be a simple linear advection equation; however, the complex dependence of $v_\Phi$ makes it highly non-linear. The flux velocity is determined by a balance between advection, which moves the magnetic flux inward toward the BH and is mediated by the large-scale flow velocities through $\mathcal{A}^\varphi$, and diffusion, which moves the flux outward and is mediated by the turbulent EMF $\mathcal{E}^\varphi$.

It is therefore natural to separate $v_\Phi$ into its advective
\begin{equation}
\label{eq:vadv_deff}
v_{\rm adv} = \frac{1}{\thmean{B^\theta}} \thmean{\mathcal{A}^\varphi}
\end{equation}
and diffusive
\begin{equation}
\label{eq:vdiff_deff}
v_{\rm diff} = \frac{1}{\thmean{B^\theta}} \thmean{\mathcal{E}^\varphi}
\end{equation}
components, where
\begin{equation}
v_\Phi = v_{\rm diff} + v_{\rm adv}.
\end{equation}

Finally, we define several useful quantities for normalizing the velocities and other parameters. For a Kerr black hole with dimensionless spin $a$, the approximate Keplerian frequency for circular orbits is
\begin{equation}
\Omega_k = \frac{1}{r^{3/2} + a},
\end{equation}
and the corresponding Keplerian velocity is
\begin{equation}
V_k = r\,\Omega_k.
\end{equation}

It is also useful to define the density scale height as
\begin{equation}
    \frac{h}{R} = \frac{\thmean{|\theta-\theta_0|\rho}}{\thmean{\rho}},
\end{equation}
where
\begin{equation}
    \theta_0 = \frac{\pi}{2}+\frac{\thmean{\left(\theta-\frac{\pi}{2}\right)\rho}}{\thmean{\rho}}.
\end{equation}
\subsection{Validation of the framework}
\label{sec:validation}

The goal of this manuscript is to develop a phenomenological, mechanism-based picture of magnetic flux variability to better understand flux eruptions in MADs. In this section, we test the formalism developed in section \ref{sec:theory} against simulation data.

\begin{figure*}
    \centering
    \includegraphics[width=\textwidth]{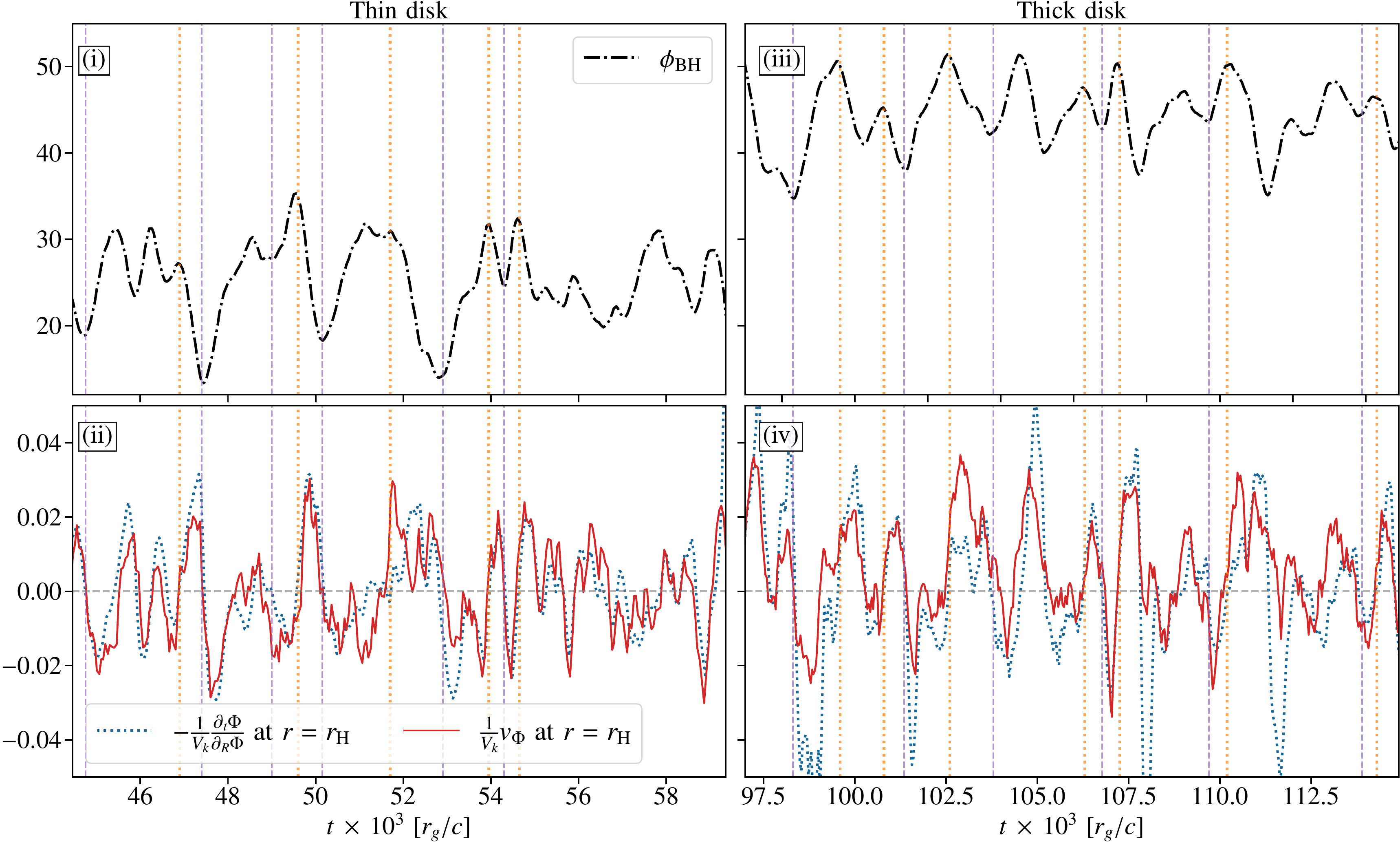}
    \caption{(i)–(iii) Evolution of $\phi_{\rm BH}$ for the thin and thick disk simulations, respectively. Dashed purple vertical lines mark local minima, and orange dotted lines mark local maxima; note the different temporal scales.
   (ii)–(iv) Comparison of $-\frac{\partial_t \Phi}{\partial_r \Phi}$ (dotted blue) and $v_\Phi$ (solid red) at $r_H$ for the thin and thick disk models, respectively. The same vertical lines as in (i)–(iii) indicate minima and maxima of $\phi_{\rm BH}$; minima are followed by inward flux advection ($v_\Phi<0$), while maxima are followed by outward flux diffusion ($v_\Phi>0$). The close agreement between the solid red and dashed blue curves supports the validity of our flux transport framework.}
    \label{fig:model_valid}
\end{figure*}

To validate our framework we  define the well known MADness parameter
\begin{equation}
    \phi_{BH} = \frac{1}{\sqrt{{\dot{m}(r=5r_g)}}}{\Phi(r=r_H)},
\end{equation}
where $\Phi$ is the radial flux defined in Eq.~\eqref{Eq:pol_flux} and used in Eq.~\ref{eq:thm_adveq}. We divide by the accretion rate
\begin{equation}
    \dot{m} = 2\pi \int\limits_0^{\pi/2} \phimean{\rho u^r} \sqrt{-g}\rm{d} \theta,
\end{equation}
evaluated at $r=5\,\,r_g$ to avoid contamination from the floors. To compute $\phi$, we first smooth $\dot{m}$ over a timescale of $100\,r_g/c$ before dividing it, removing fast variations in $\dot{m}$ while still accounting for the long-term mass loss in the system.
We show the MADness parameter for both thin and thick disks in Fig.~\eqref{fig:model_valid}(i) and (iii), respectively. The dashed purple vertical lines indicate the local minima of $\phi_{\rm BH}$, while the orange dotted lines mark the local maxima of $\phi_{\rm BH}$.
We note that both simulations are not run for the same duration.

Clear flux eruptions are visible in both models, appearing as large oscillations in the magnetic flux attached to the black hole. Both exhibit a similar recurrence timescale for large eruptions, $t_{\rm rec} \sim1000- 2500\,\,r_g/c$. 
The average magnitude of the flux differs between the two cases. The thin disk model shows an average ${\phi}_{\rm BH} \sim 25$, while the thick disk model averages to ${\phi}_{\rm BH} \sim 45$.
The average value of $\phi_{\rm BH}$ and the observed recurrence timescale for both thin and thick disks is consistent with previous results in the literature \citep{tchekhovskoy_efficient_2011,avara_efficiency_2016,scepi_magnetic_2023}.

To confirm the validity of the formalism developed in Section \ref{sec:theory}, we start by defining an alternative flux transport velocity.  From Eq.~\eqref{eq:thm_adveq} one can write
\begin{equation}
    v_\Phi = -\frac{\partial_t \Phi}{\partial_r \Phi},
\end{equation}
which together with the definition of $v_{\Phi}$ from Eq.~\ref{eq:vphi_defs} can be used to verify the validity of  Eq.~\eqref{eq:thm_adveq}.

In Fig.~\ref{fig:model_valid} (ii) and (iv), we show both $-\frac{\partial_t \Phi}{\partial_r \Phi}$ (dashed blue line) and $v_\Phi$ (solid red line), evaluated at $r_H$, as functions of time for the thin (ii) and thick (iv) disk models. We observe good agreement between the two curves for both disk models. Computing the correlation between both curves yields values around $\sim 0.7$ for both models, indicating a strong correlation and providing solid validation of the magnetic field transport framework developed in Section~\ref{sec:theory}.

The dashed purple vertical lines in Fig.~\ref{fig:model_valid} (ii) and (iv) indicate local minima of $\phi_{\rm BH}$, while the orange dotted lines mark local maxima. We find that minima in $\phi_{\rm BH}$ are followed by inward flux advection ($v_\Phi < 0$), while maxima are followed by outward flux diffusion ($v_\Phi > 0$). As expected, the flux transport velocity $v_\Phi$ and $\phi_{\rm BH}$ reflect each other’s evolution, confirming their dynamic coupling.

\begin{figure*}
    \centering
    \includegraphics[width=\textwidth]{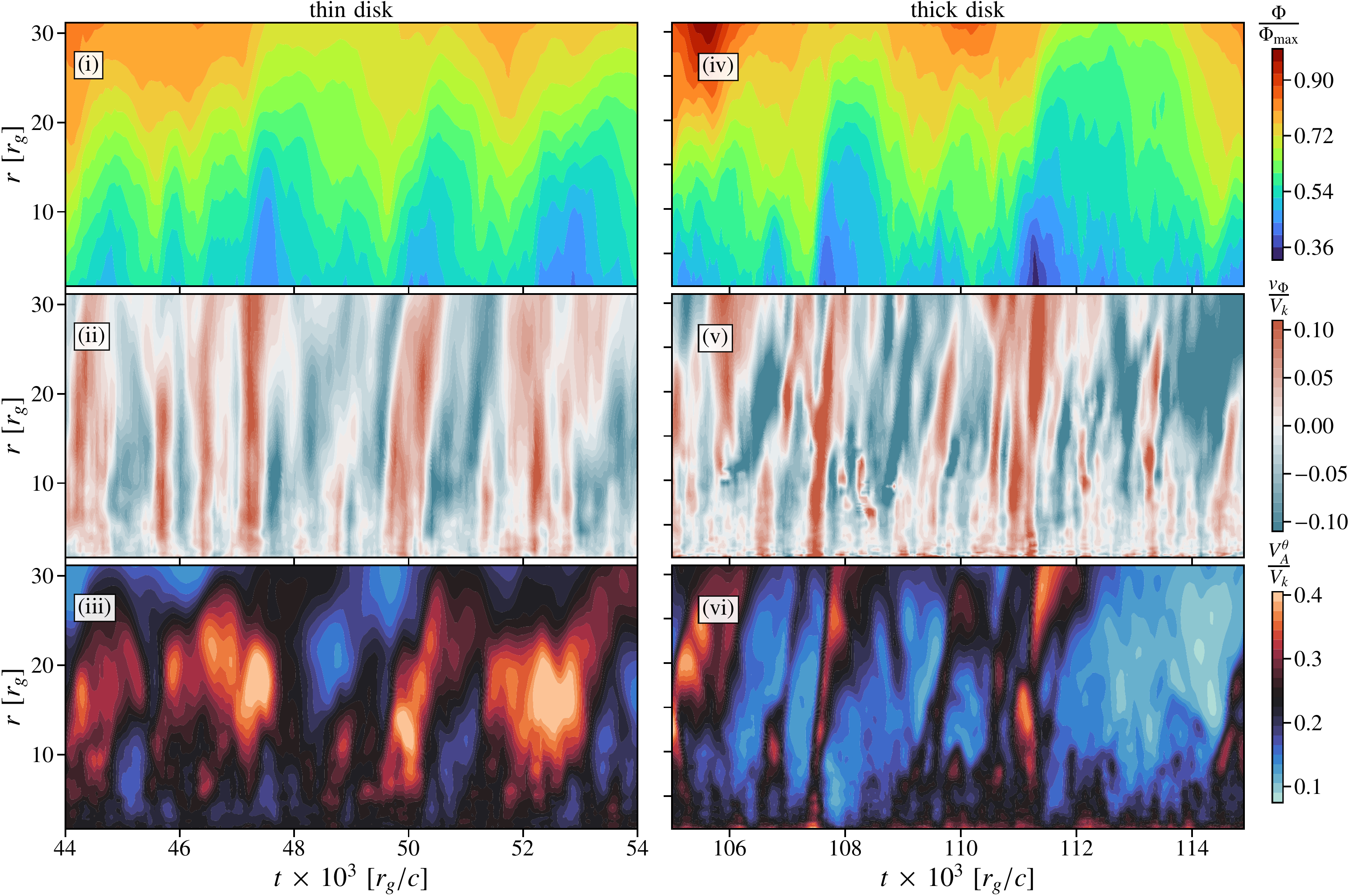}
    \caption{(i)–(iv) $r$–$t$ spacetime diagrams of the magnetic flux $\Phi(r,t)$ for the thin and thick disk models, respectively. Contours trace magnetic field lines, revealing cyclical inward and outward transport related to, but distinct from, flux eruptions (see text).
(ii)–(v) $r$–$t$ spacetime diagrams of the flux transport velocity $v_\Phi(r,t)$ for the thin and thick disks. Red and blue regions correspond to inward advection and outward diffusion, respectively, and their patterns closely match those seen in $\Phi$ panels (i)–(iv), indicating consistency between flux evolution and transport velocity.
(iii)–(vi) $r$–$t$ spacetime diagrams of the Alfvén velocity associated with the vertical magnetic field, $V_A^\theta/V_K$. Flux eruptions appear as regions with elevated $V_A^\theta/V_K$, with a clear temporal correlation between these high-$V_A^\theta/V_K$ events and the diffusion phases identified in panels (iii)–(iv). This agreement further supports the validity of our flux transport framework.}
    \label{fig:r_t_erupt}
\end{figure*}

    
Figure~\ref{fig:r_t_erupt} shows the $r$–$t$ spacetime diagram of the magnetic flux $\Phi(r,t)$ for both thin (i) and thick (iv) disk models. Each contour represents the position of a magnetic field line, allowing us to track its transport through the disk. We observe cyclical inward and outward motions of the magnetic field. 

While it is tempting to associate all outward transport of magnetic flux with the violent, non-axisymmetric flux eruptions that propagate outward as localized bubbles of enhanced magnetic field strength, in this work we make a more nuanced distinction. We refer to the outward radial transport of magnetic flux as magnetic diffusion phases. These are correlated with, but distinct from, flux eruptions—they lack the non-axisymmetric character and instead represent axisymmetric transport processes.

We show in Fig.~\ref{fig:model_valid}(i,iv) flux diffusion events, where the magnetic field propagates radially outward, reaching radii of up to $r \sim 30\,r_g$ for the thin disk and $r \sim 60\,r_g$ for the thick disk (not shown). The maximum radial extent of these diffusion phases likely depends on both the resolution and the radial range over which the system has reached steady state. After each diffusion phases, the flux is advected back inward, replenishing the magnetic flux at the horizon.

In Fig.~\ref{fig:r_t_erupt}, we also show the $r$–$t$ spacetime diagram of the flux transport velocity, $v_\Phi(r,t)$, for both the thin (ii) and thick (v) disk models. Red and blue stripes correspond to periods of advection and diffusion of the large-scale magnetic field. These patterns closely match those seen in the magnetic flux $\Phi$ panels (i) and (iv), demonstrating consistency between the flux and transport velocity. This agreement confirms that our formalism captures magnetic field transport reliably across all radii far from the black hole, provided the system has reached a statistical quasi-steady state.

Finally, Figure~\ref{fig:r_t_erupt} shows the $r$–$t$ spacetime diagram of the Alfvén velocity associated with the vertical magnetic field, computed as 
\begin{equation}
V_A^\theta = \left. \sqrt{\frac{\phimean{b^\theta b_\theta}}{\phimean{\rho}}}\right|_{\theta = \pi/2}
\end{equation}
and normalized to the local Keplerian velocity,  We plot $V_{A}^\theta/V_K$ for both the thin (iii) and thick (vi) disk models. 

To preserve non-axisymmetric features, the square is taken before $\varphi$-averaging, allowing this quantity to better trace the location of flux eruptions. Flux eruptions are visible as regions with elevated values of $V_{A}^\theta/V_K$, their shape is consistent with features identified in previous studies \citep{porth_flares_2021}. Notably, these eruptions persist longer in the thin disk (iii) than in the thick disk (vi).

We also observe a temporal correlation between the flux eruptions identified via $V_{A}^\theta/V_K$ in Fig.\ref{fig:r_t_erupt} (iii) and (vi) and the flux diffusion events. The latter are tracked using the large-scale magnetic flux, $\Phi$ (Fig.\ref{fig:r_t_erupt} (i) and (iv)), and the flux advection velocity, $v_\Phi/V_K$ (Fig.~\ref{fig:r_t_erupt} (ii) and (v)).  However, this correlation is not exact: we find clear differences in both the duration and radial structure of these events. We conclude that although flux eruptions and flux diffusion events are dynamically related, they are not necessarily the same physical process. Flux diffusion tracks the evolution of the global, large-scale magnetic field, whereas flux eruptions also involve contributions from small-scale turbulence within the accretion flow.

\section{Flux transport in magnetically arrested disks}
\label{sec:flux_trans_results}
\subsection{Phenomenology of flux transport}
\begin{figure*}
    \centering
    \includegraphics[width=\textwidth]{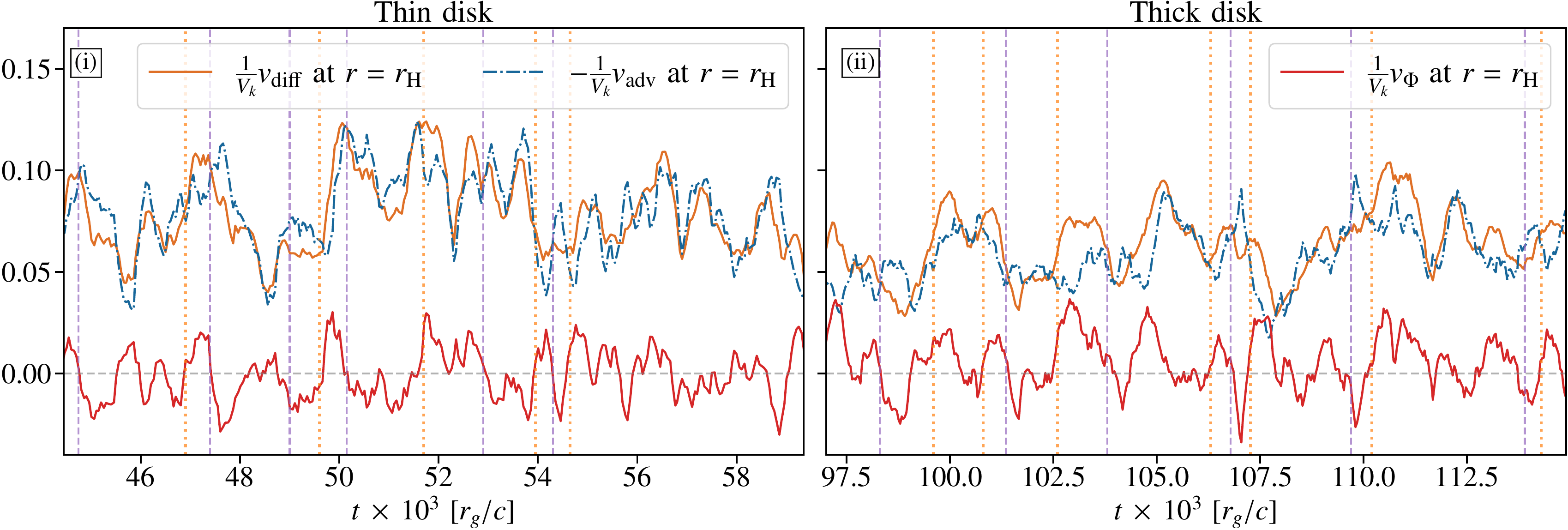}
    \caption{Time evolution of $v_{\rm adv}$, $v_{\rm diff}$, and their sum $v_\Phi$ at $r = r_H$, normalized to the local Keplerian velocity $V_k$, for the thin (i) and thick (ii) disk models. In both cases, $v_{\rm adv} < 0$ and $v_{\rm diff} > 0$, showing inward advection and outward diffusion as expected, supporting the validity of our framework. We also observe $v_{\rm diff} \simeq |v_{\rm adv}| \gg |v_\Phi|$, indicating that net flux transport arises from small deviations between advection and diffusion. On average, $v_\Phi \sim 0$, consistent with a statistical steady state of the magnetic structure, though strong fluctuations remain.  Dashed purple vertical lines mark local minima, and orange dotted lines mark local maxima; obtained from Fig.~\ref{fig:model_valid}.}
    \label{fig:vdiff_vadv}
\end{figure*}

Using Eqs.~(\ref{eq:vadv_deff},~\ref{eq:vdiff_deff}), we define the advective ($v_{\rm adv}$) and diffusive ($v_{\rm diff}$) components of the flux transport velocity, $v_\Phi$. Figure~\ref{fig:vdiff_vadv} shows the time evolution of $v_{\rm adv}$, $v_{\rm diff}$, and their sum $v_\Phi$ at $r=r_H$, all normalized to local $V_k$, for both the thin (i) and thick (ii) disk models. In both cases, we observe that $v_{\rm diff} > 0$ and $v_{\rm adv} < 0$. This clear distinction—outward diffusion and inward advection behaving as expected—is both physically consistent and reassuring for the validity of our framework. 
We also find that $|v_{\rm diff}|\simeq |v_{\rm adv}|$ and $|v_{\rm diff}|,|v_{\rm adv}|\gg |v_\Phi|$ for both thin and thick disk models. This implies that any measurable flux transport is the result of small differences between the advection and diffusion speeds. 
In this view the variability described above including flux eruptions, and related to diffusion  and advection phases, represents small deviations in the overall equilibrium of the larger advective and diffusive fluxes of magnetic field.  The magnetic structure has reached a statistical steady state, with $|v_\Phi|\sim 0$ on average (see averages next Section), and we interpret diffusion (and their corresponding flux eruptions) and advection phases as fluctuations around this statistical steady state. 


This interpretation of flux eruptions as emerging from small differences between advection and diffusion would seem to contradict the current literature \citep{porth_flares_2021,ripperda_black_2022}. Flux eruptions are usually interpreted as reconnection events in the black hole magnetosphere that travel radially outward through the disk, rather than processes primarily related to disks physics. In reality, there is no contradiction between our framework and the current understanding of flux eruptions. Indeed, our framework just provides a means for  measuring how the disk responds to flux eruptions, which originate in the magnetosphere but enhance the turbulent dissipation within the disk as they propagate outward (see section \ref{sec:eff_res}).

{We also measured the advective ($v_{\rm adv}$) and diffusive ($v_{\rm diff}$) terms at early times (not shown), before the disk becomes MAD ($t\lesssim3000\,r_g/c$). The system first undergoes strong relaxation transients as the disk forms from the initial torus and matter falls onto the black hole ($t\lesssim1200\,r_g/c$), during which velocities are hard to interpret. It then settles into an advection phase ($1200\lesssim t\lesssim2000\,r_g/c$) where $|v_{\rm adv}|\gg |v_{\rm diff}|$ and $v_\Phi\simeq v_{\rm adv}\simeq-0.1V_k$ at $r=r_H$. For $t\gtrsim2000\,r_g/c$, $v_{\rm diff}$ increases, reducing the flux transport velocity as the system saturates into the MAD state around $t\simeq3000\,r_g/c$ with $|v_\Phi|\ll |v_{\rm adv}|\simeq |v_{\rm diff}|$. Thus, the MAD state can be interpreted as a magnetic steady state achieved through enhanced diffusion, with equilibrium defined by $|v_{\rm adv}|\simeq |v_{\rm diff}|$.  

Flux eruptions appear then as discrete events triggered by the accumulation of magnetic flux. Once sufficient flux builds up, it erupts, causing large-scale diffusion of the magnetic field as the turbulent diffusivity temporarily exceeds advection (see section \ref{sec:eff_res}). After the diffusive phase, the system returns to an advective phase, where diffusion is slightly smaller than advection, allowing inward transport and flux accumulation around the black hole, restarting the cycle.}


\subsection{Radial and vertical structure}
\label{sec:radvert_struc}
We now examine the radial structure of the magnetic field transport velocities by averaging them over time. This allows us to better understand where magnetic field transport is concentrated both radially and vertically, assess the role of vertical stratification, and determine how the efficiency of magnetic field transport varies with radius. This will be leveraged in Section~\ref{sec:trec} to construct a model for the recurrence timescale of flux eruptions.

\begin{figure}
    \centering
    \includegraphics[width=\columnwidth]{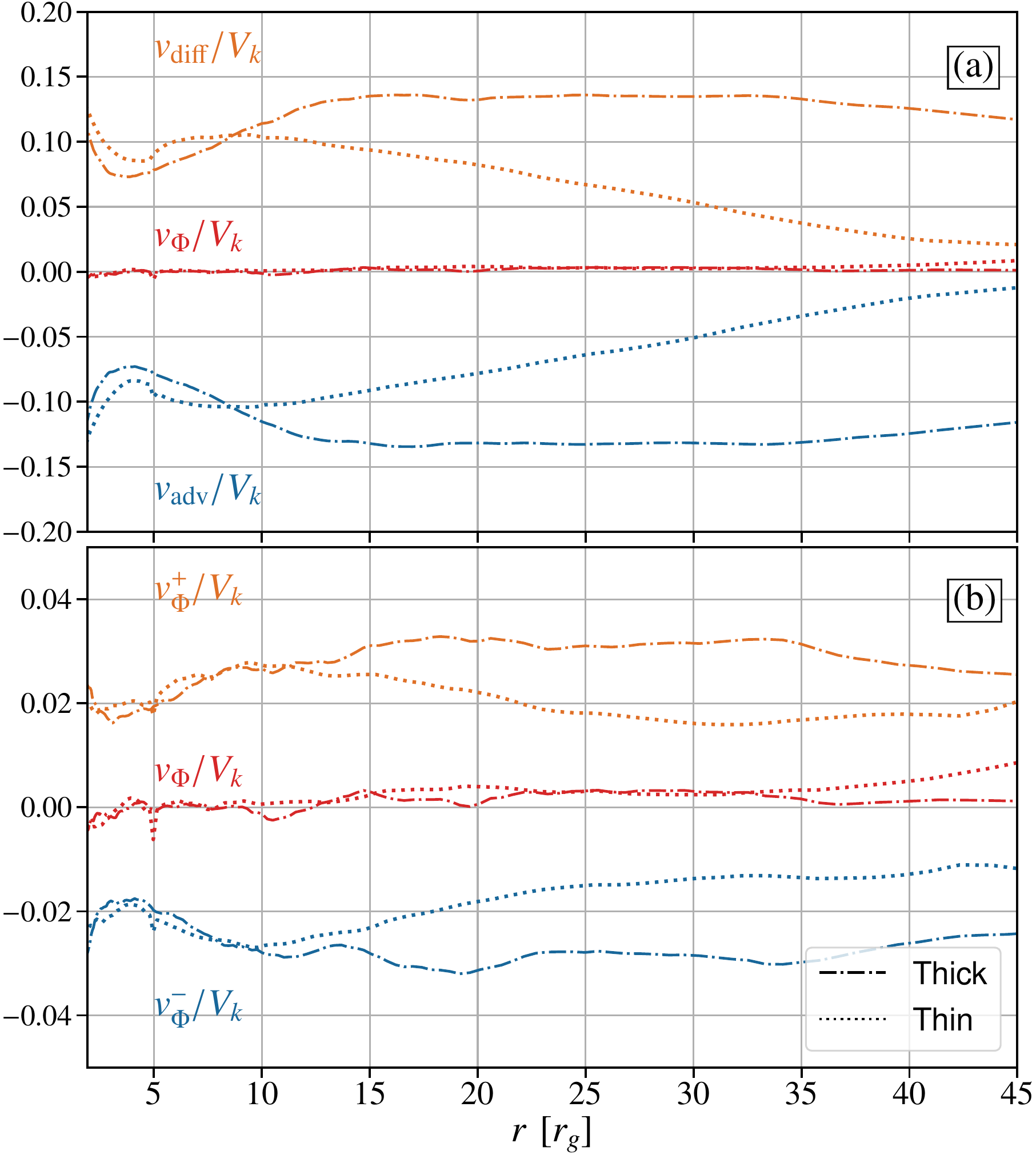}
    \caption{(a) Time-averaged radial profiles of $v_\Phi$, $v_{\rm adv}$, and $v_{\rm diff}$, normalized to the local Keplerian velocity, for both disk models. Across all radii, $v_{\rm adv} < 0$ and $v_{\rm diff} > 0$, with magnitudes much larger than $v_\Phi$, showing that net flux transport arises from a small difference between advection and diffusion. The profiles of $v_{\rm adv}/V_k$ and $v_{\rm diff}/V_k$ are relatively flat, suggesting that in a fully steady-state simulation, these profiles remain flat for $r \gtrsim 10\,r_g$.
    (b) Since $v_\Phi$ is stochastic, its time average tends to $v_\Phi \sim 0$. To better illustrate its structure, the positive and negative contributions, $v_\Phi^+$ and $v_\Phi^-$, are separately time-averaged and normalized to $V_k$, these are also called the net outward and inward velocities respectively. They are smaller than $v_{\rm adv}$ and $v_{\rm diff}$ by a factor of $\sim 3$–$5$, indicating that the residual net flux transport is much smaller than the total advection and diffusion contributions. The profiles of $v_\Phi^{+,-}/V_k$ are roughly constant over $r_H \lesssim r \lesssim 40\,\,r_g$.}
    \label{fig:rad_struc_v}
\end{figure}

Figure~\ref{fig:rad_struc_v}(a) shows the $t$-averaged radial profiles of $v_\Phi$, $v_{\rm adv}$, and $v_{\rm diff}$, each normalized to the local Keplerian velocity. Across all radii and for both disk models, we consistently find that time average $v_{\rm diff} > 0$ and $v_{\rm adv} < 0$, with both having much larger magnitudes than $v_\Phi$. This reinforces our conclusion that $v_\Phi$ arises as a small residual between the opposing effects of advection and diffusion. Additionally, we confirm that the time-averaged $v_\Phi \approx 0$, indicating the system is in a statistically quasi-steady state. However, we observe that this clean separation between advection and diffusion begins to break down at larger radii ($r \gtrsim 25\,\,r_g$) in the thin disk model. We attribute this to those outer regions not yet being in full steady state—expected given that the thin disk was evolved for only $\sim 60,000\,r_g/c$, compared to $\sim 120,000\,r_g/c$ for the thick disk—and that thinner disks typically require longer times to reach equilibrium. We note that the profile of the advective and diffusive velocities for $r\lesssim8\,\,r_g$ are remarkably similar. 
Finally, we note that the profiles of $v_{\rm diff}/V_k$ and $v_{\rm adv}/V_k$ are relatively flat in specific radial ranges, $5\,r_g < r < 15\,r_g$ for the thin disk and $15\,r_g < r < 40\,r_g$ for the thick disk. We speculate that in a simulation that has reached statistical steady state out to large radii, $v_{\rm diff}/V_k$ and $v_{\rm adv}/V_k$ should exhibit flat profiles for $r\gtrsim 10\,r_g$. We note that this radial dependency likely results from the slowly varying $h/R$ in the thick disk and the constant $h/R$ in the thin disk. Disks with significant radial variations in $h/R$ would exhibit different radial dependencies of the magnetic flux velocities.

As shown above, $v_\Phi$ is a stochastic quantity and therefore averaging it in time will lead to $v_\Phi \sim 0$. To represent this quantity more clearly, we time-average its positive and negative contributions separately, allowing us to assess whether the radial and vertical structure changes significantly during advective ($v_\Phi < 0$) or diffusive ($v_\Phi > 0$) phases of field transport.

This is shown in Figure~\ref{fig:rad_struc_v}(b), where $v_\Phi^+$ represents the time averaged positive contributions of $v_\Phi$ and $v_\Phi^-$ represents the time averaged negative contributions of $v_\Phi$ both normalized to $V_k$. We find that $v_\Phi^{+,-}$ are smaller than $v_{\rm adv,diff}$ by a factor of $\sim 3$–$5$. This indicates that the residual net transport of magnetic field is much smaller than the total contributions from advection and diffusion. Additionally, the profiles of $v_\Phi^{+,-}/V_k$ are roughly constant with radius over the range $r_H \gtrsim r \gtrsim 40\,\,r_g$.

Interestingly, we measure a small but nonzero positive total flux velocity for $r \gtrsim 10\,\,r_g$, with $v_\Phi \sim 7 \times 10^{-3}$ for both thin and thick disk models. This signal is naturally interpreted as the slow redistribution of large-scale magnetic flux toward larger radii. Indeed, it is well established that MADs form from the inside out, with strong large-scale fields gradually spreading to larger radii over the course of the system's evolution.

\begin{figure}
    \centering
    \includegraphics[width=\columnwidth]{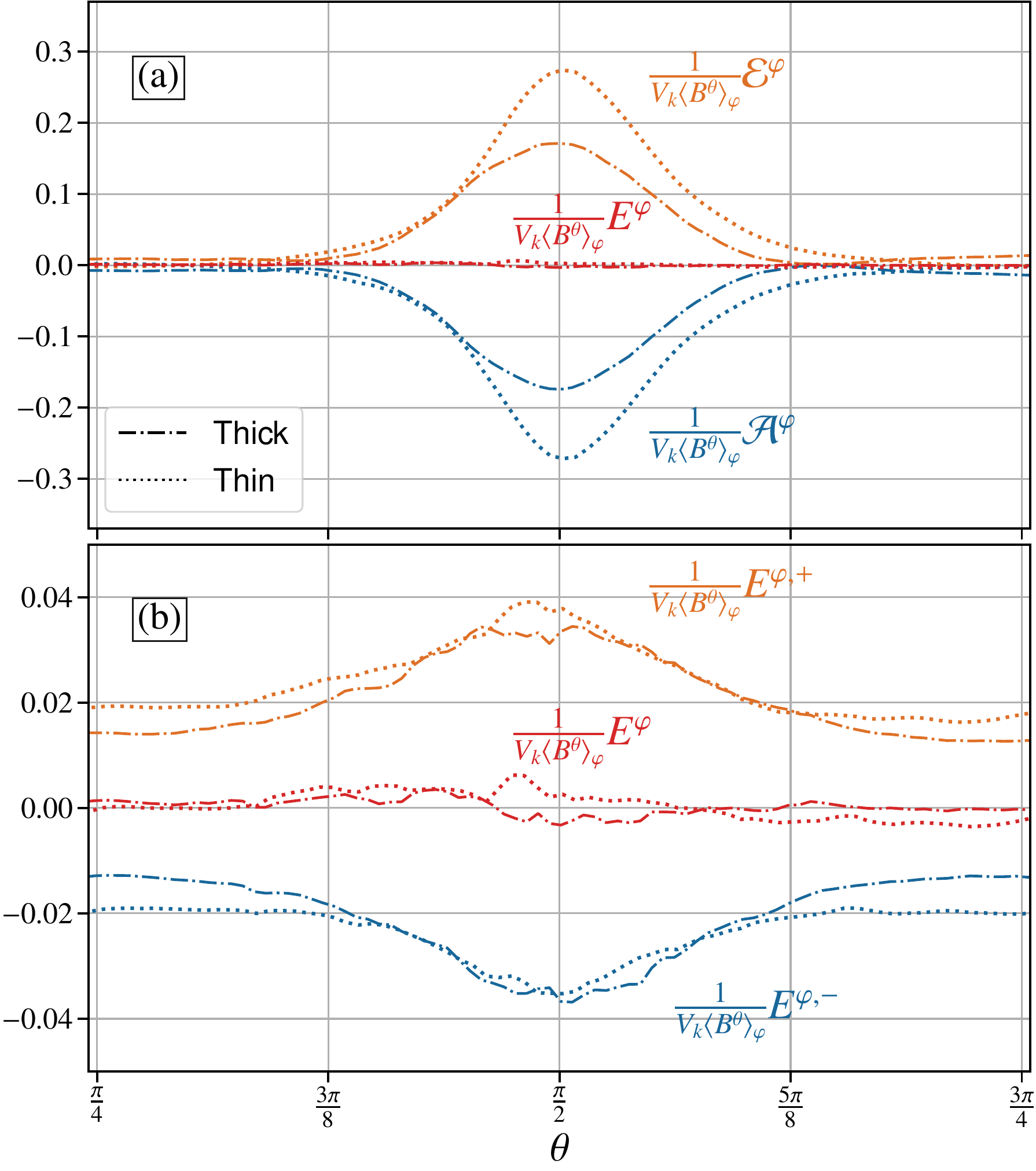}
    \caption{Vertical profiles of flux transport velocities at $r=6\,r_g$. Panel (a): $\mathcal{A}^\varphi/\langle B^\theta\rangle_\varphi$ (advection velocity $v_{\rm adv}$), $\mathcal{E}^\varphi/\langle B^\theta\rangle_\varphi$ (diffusion velocity $v_{\rm diff}$), and $E^\varphi/\langle B^\theta\rangle_\varphi$ (total flux transport velocity $v_\Phi$), all normalized to the local Keplerian velocity. Both thin and thick disk models show Gaussian-like profiles for $\mathcal{A}^\varphi/\langle B^\theta\rangle_\varphi$ and $\mathcal{E}^\varphi/\langle B^\theta\rangle_\varphi$, while the time-averaged $E^\varphi/\langle B^\theta\rangle_\varphi$ remains close to zero. Thick disks display a slightly shallower vertical decrease compared to thin disks, as expected. Panel (b): latitudinal profiles of the positive ($E^{\varphi,+}/\langle B^\theta\rangle_\varphi$) and negative ($E^{\varphi,-}/\langle B^\theta\rangle_\varphi$) contributions to the magnetic flux velocity, alongside the net $E^{\varphi}/\langle B^\theta\rangle_\varphi$. The positive and negative contributions are each an order of magnitude smaller than the advection and diffusion terms, consistent with the near cancellation in the total transport velocity.}
    \label{fig:vert_struc_v}
\end{figure}

Now we focus on the $t$-averaged latitudinal structures of the advection and diffusion velocities. In Fig.~\ref{fig:vert_struc_v} (a) we show the vertical structure of $\mathcal{A}^\varphi/\langle B^\theta\rangle_\varphi$, $\mathcal{E}^\varphi/\langle B^\theta\rangle_\varphi$, and ${E}^\varphi/\langle B^\theta\rangle_\varphi$ normalized to the local Keplerian velocity and evaluated at $r=6\,r_g$. We remind the reader that $\mathcal{A}^\varphi/\langle B^\theta\rangle_\varphi$ is equivalent to the advective velocity $v_{\rm adv}$, $\mathcal{E}^\varphi/\langle B^\theta\rangle_\varphi$ is equivalent to the diffusive velocity $v_{\rm diff}$ and $E^\varphi/\langle B^\theta\rangle_\varphi$ is equivalent to the total flux transport velocity $v_{\Phi}$. The different notation is to differentiate them from their vertically average counterparts. 

We distinguish Gaussian-like profiles for the advection and diffusion velocities, while again we find $t$-average of the net flux velocity $E^\varphi/\langle B^\theta\rangle_\varphi \simeq 0$. It is interesting that thin and thick disk models show very similar latitudinal profiles, with the thick disk having a slightly shallower vertical decrease, as would be expected due to larger $h/R$. Gaussian profiles are also encouraging as they might be easier to model analytically. Furthermore, this verifies our approximation that most of the magnetic field transport is concentrated near the disk midplane.

In Fig.~\ref{fig:vert_struc_v} (b) we show the latitudinal profiles of the positive $E^{\varphi,+}/\langle B^\theta\rangle_\varphi$ and negative $E^{\varphi,-}/\langle B^\theta\rangle_\varphi$ contributions to the magnetic flux velocity and the net magnetic flux velocity $E^\varphi/\langle B^\theta\rangle_\varphi$, where we have again separately time-averaged positive and negative contribution of the net magnetic flux velocity.
We notice again that the net outward and inward velocities $E^{\varphi,+}/\langle B^\theta\rangle_\varphi$ and $E^{\varphi,-}/\langle B^\theta\rangle_\varphi$ are an order of magnitude smaller than the advection and diffusion terms.

\subsection{Recovering the recurrence timescale: a toy model for flux eruptions}
\label{sec:trec}
\begin{figure*}
    \centering
    \includegraphics[width=0.9\textwidth]{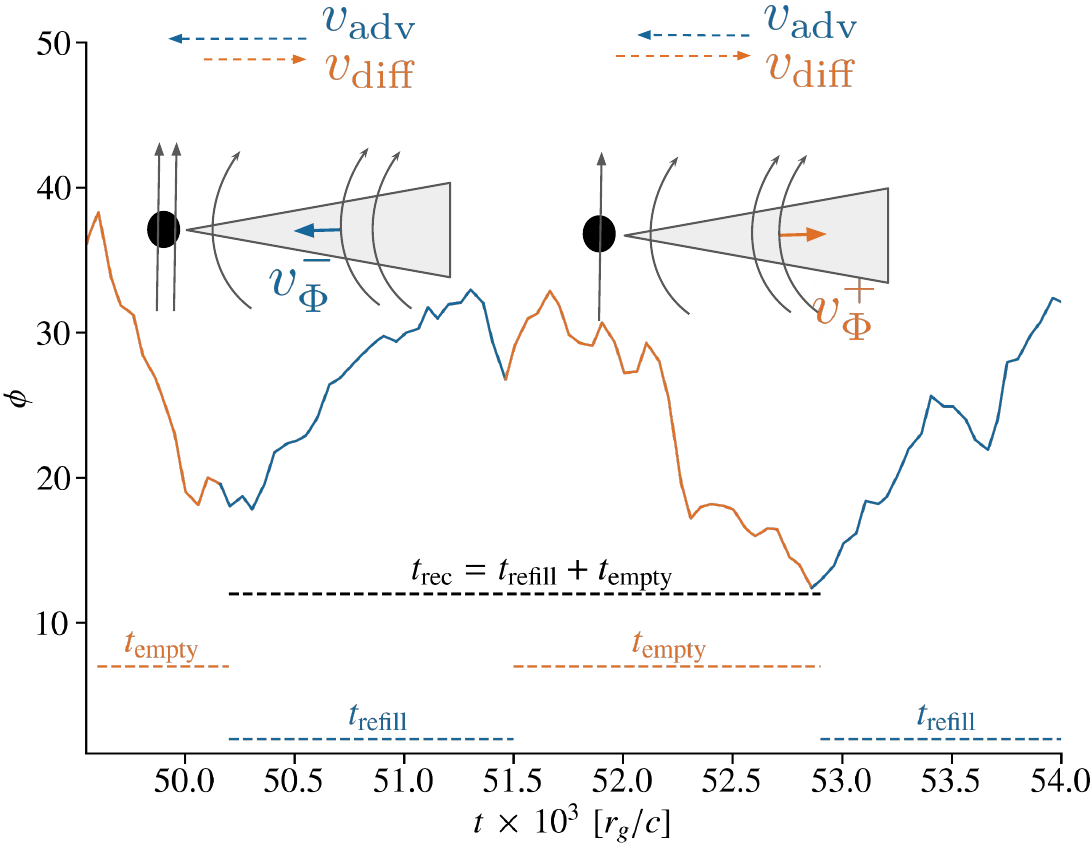}
    \caption{Summary of the physical framework and results. The flux, $\phi$, is shown during an eruption to illustrate the mechanisms driving flux transport. Episodes of inward net transport ($v_\phi^-$) and outward net transport ($v_\phi^+$) emerge from the balance between advection ($v_{\rm adv}$) and diffusion ($v_{\rm diff}$). These processes cause the flux to empty ($t_{\rm empty}$) during diffusion or refill ($t_{\rm refill}$) during advection, with the recurrence timescale of flux eruptions given by $t_{\rm rec} = t_{\rm empty} + t_{\rm refill}$.}
    \label{fig:diagram}
\end{figure*}

Magnetically arrested disks undergo flux eruptions approximately every $t_{\rm rec} \sim 1000\,\,r_g/c$, with similar recurrence timescales for both thin and thick disks \citep{tchekhovskoy_efficient_2011,avara_efficiency_2016,scepi_magnetic_2023}. In this work, we interpret these eruptions as part of magnetic flux diffusion episodes within the disk, followed by phases of inward advection.
Building on this understanding, we derive an analytical expression for the recurrence timescale based on the radial structure of the magnetic flux $\phi$, the flux transport velocity $v_\Phi$, and the flux change across an eruption, $\delta \phi_{\rm BH}$, which we estimate as twice the standard deviation of $\phi_{\rm BH}(t)$ measured on the BH horizon.

Our physical framework and results are summarized in Fig.~\ref{fig:diagram}, where we zoom in on the flux, $\phi$, during an eruption to illustrate the mechanisms driving flux transport. We show how episodes of inward net transport ($v_\phi^-$) and outward net transport ($v_\phi^+$) arise from the balance between advection ($v_{\rm adv}$) and diffusion ($v_{\rm diff}$). These processes lead to episodes where the flux empties ($t_{\rm empty}$) during diffusion or refills ($t_{\rm refill}$) during advection. The recurrence timescale of flux eruptions, $t_{\rm rec}$, is then simply the sum of these two timescales.

One would expect the magnetic field to be advected at a velocity $|v_{\rm adv}|\sim |v_{\rm acc}|$, where the accretion velocity is found to be $|v_{\rm acc}|\sim0.2\,V_k$ \citep{scepi_magnetic_2023}, consistent with our result of $|v_{\rm adv}|\sim0.2\,V_k$. However, as shown above, the advection is balanced by the diffusive velocity, $|v_{\rm diff}|\sim0.2\,V_k$, which in turn leads to a much smaller flux transport velocity, by about an order of magnitude, $|v_{\Phi}|\sim10^{-2}\,V_k$. This helps explain why the recurrence timescales of MAD flux eruptions are so long, since the magnetic field is transported at a much smaller velocity than one would naively expect from estimates based only on the accretion velocity.

From Section~\ref{sec:radvert_struc}, we conclude that the large-scale magnetic field is transported inward at a velocity\footnote{We note that this may be related to the slowly varying $h/R$ in the thick disk and the constant $h/R$ in the thin disk. Disks with stronger $h/R$ variations would exhibit different radial dependencies in magnetic flux velocities.} $v_\Phi^{-} \propto V_k$. A similar conclusion holds for the outward transport velocity $v_\Phi^{+}$. To model this behavior, we choose the following form for the velocities
\begin{equation}
v_\Phi^{\pm}(r) = \pm v_0\Tfrac{r}{r_g}^{-1/2},
\end{equation}
where $v_0 \simeq 1.3$–$3.5 \times 10^{-2}$. This choice is consistent with the data shown in Fig.~\ref{fig:rad_struc_v}, while also allowing for an analytical solution below.

It is clear that the recurrence timescale exceeds what would be expected from a simple estimate based on the inward net flux velocity,
\begin{equation}
    t_{ \phi }^{-} = \frac{r}{v_\phi^{-}}\simeq 30-80\,\, \Tfrac{r}{r_g}^{3/2}\,\,r_g/c.
\end{equation}
However, $t_\phi^{-}$ shows a steep radial dependence, scaling as $\propto r^{3/2}$, which implies that large-scale field transport becomes increasingly inefficient at larger distances from the central BH. At the same time, the large-scale flux $\phi$ tends to increase with radius, making more flux available farther out, which could partially counteract the decline in transport efficiency. A reliable prediction of the flux eruption recurrence timescale $t_{\rm rec}$ therefore requires consideration of both radial profiles.

Our computation of $t_{\rm rec}$ is grounded on a solution of the advection equation used throughout this paper,
\begin{equation}
\partial_t\phi_{\rm th} + v_\Phi^{-}\partial_r\phi_{\rm th} = 0,
\label{eq:adv_trec}
\end{equation}
where we adopt the net inward velocity $v_\Phi^{-}$ as the transport speed. For consistency with the literature, we use $\phi_{\rm th} = \Phi_{\rm th}/\sqrt{\dot{m}}$, assuming $\dot{m}$ to be constant in this calculation. The subscript $_{\rm th}$ distinguishes our theoretical solution of $\phi_{\rm th}$ from the numerical results discussed elsewhere in the manuscript.

To solve Eq.~\eqref{eq:adv_trec}, we first require an initial condition for the radial flux. We model the average flux with
\begin{equation}
\phi_{\rm ini} = \phi_{\rm BH}^{\rm th} + A\Tfrac{r}{r_g}^{3/4},
\end{equation}
where $\phi_{\rm BH}^{\rm th}$ is the average flux anchored at the BH event horizon\footnote{\label{foot:phibh_val}The exact horizon flux is $\phi_{\rm BH}^{\rm th} + A\Tfrac{r_H}{r_g}^{3/4}$, but the term $A\Tfrac{r_H}{r_g}^{3/4}$ is smaller by an order of magnitude.}. The exponent $3/4$ is adopted from the \cite{blandford_hydromagnetic_1982} theory, while  $A$ is a constant of order unity chosen to match the radial profile of the large-scale flux.

\begin{figure*}
    \centering
    \includegraphics[width=0.9\textwidth]{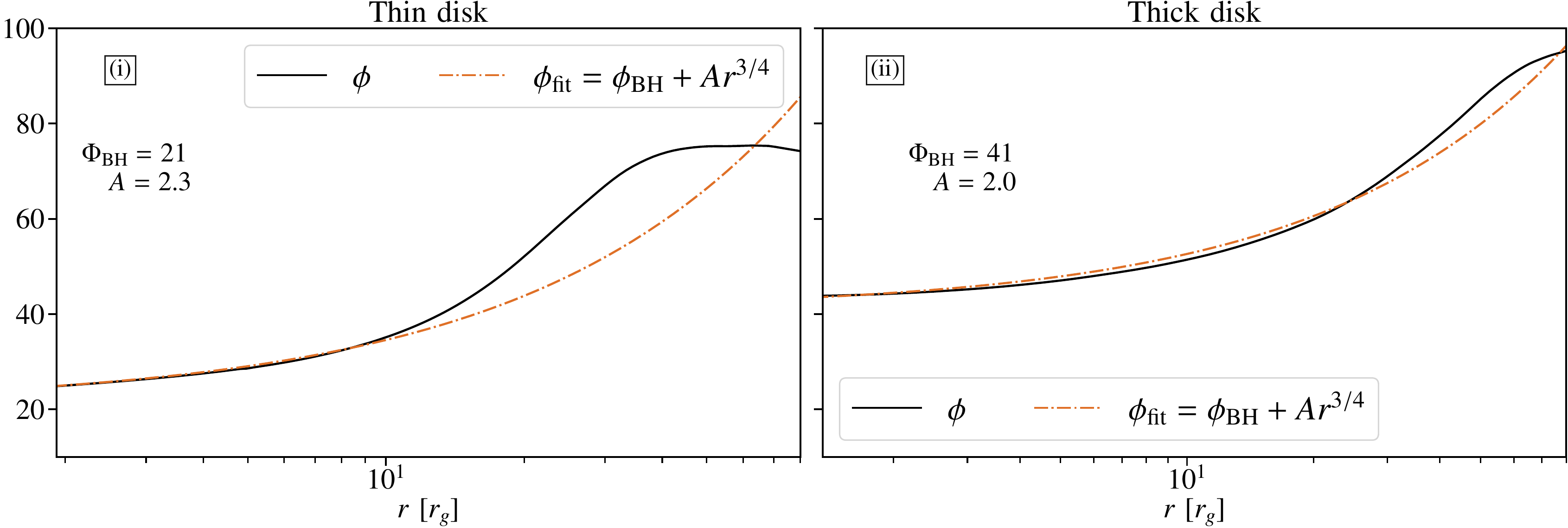}
    \caption{Comparison of the time-averaged dimensionless magnetic flux, $\phi$, and its fitted profile as functions of radius for the thin (i) and thick (ii) disk simulations. The fit is restricted to the inner regions ($r < 15\,r_g$), where the simulations have numerically converged, particularly for the thin disk (see Section~\ref{sec:radvert_struc}). Both disk models are well fit with a similar parameter, $A \simeq 2$. The flux at the horizon differs between the two disks, with $\phi_{\rm BH} = 21$ for the thin disk and $\phi_{\rm BH} = 41$ for the thick disk.}
    \label{fig:fit_phi}
\end{figure*}

In Figure~\ref{fig:fit_phi}, we compare the average\footnote{where we t-average $\Phi(r=r_{H},t)$ and $\dot{m}(r=5\,\,r_g,t)$ before computing $\phi$.} $\phi$ and its fitted profile as functions of radius for the thin (i) and thick (ii) disk simulations. We restrict our fit to the inner regions ($r<15\,\,r_g$), as those are the zones where the simulations have numerically converged—particularly important for the thin disk, as discussed in Section~\ref{sec:radvert_struc}.

We find that both thin and thick disks can be fit using very similar values for $A$,  i.e., $A\simeq 2$ and different values of $\phi_{\rm BH}$. As shown throughout the manuscript, the flux at the horizon differs by a factor of approximately $2$, with $\phi_{\rm BH}^{\rm th}=21$ for the thin disk and $\phi_{\rm BH}^{\rm th}=41$ for the thick disk\footnote{The real values of $\phi_{\rm ini}(r=r_{\rm{H}})$ are slightly larger, being $24$ and $43$ respectively see footnote \ref{foot:phibh_val}}.

Now, armed with an initial condition $\phi_{\rm ini}$ and a formula for $v_\phi^-$, we can solve Eq.~\eqref{eq:adv_trec}. This equation is solved using the method of characteristics to obtain
\begin{equation}
\phi_{\rm th}(r,t) = A\left(\frac{3}{2}\frac{v_0}{r_g}t+\Tfrac{r}{r_g}^{3/2}\right)^{\frac{1}{2}}+\phi_{\rm BH}^{\rm th}.
\end{equation}
This closed-form solution for the field's evolution allows us to address the following question: after a flux eruption, how long does it take for the field to refill the BH to its original value, enabling another flux eruption? This can be formally expressed as
\begin{equation}
\label{eq:solve_for_t_refill}
\phi_{\rm th}(r_{\rm{H}},t_{\rm refill}) = \phi_{\rm BH}^{\rm th} +\delta \phi,
\end{equation}
where $\delta\phi$ is the average amount of flux that must be advected to trigger a new eruption.
For simplicity, we set $\delta \phi = 2\sigma_\phi$, where $\sigma_\phi$ is the standard deviation of the time signal $\phi_{\rm BH}(t)$ shown in Fig.~\ref{fig:model_valid}. We find this quantity to be $\sigma_\phi\simeq 5$ for both thin and thick disk models. Solving Eq.~\ref{eq:solve_for_t_refill}, we obtain
\begin{align}
t_{\rm refill} &= \frac{2}{3}\frac{r_g}{v_0}\left[\left(\frac{\delta\phi}{A}\right)^{2}-\Tfrac{r_H}{r_g}^{3/2}\right]\\
t_{\rm refill} &\simeq \frac{2}{3}\frac{r_g}{v_0}\left(\frac{\delta\phi}{A}\right)^{2}=\frac{2}{3}\frac{r_g}{v_0}\sigma_\phi^{2},
\end{align}
where we safely neglect the term $\Tfrac{r_H}{r_g}^{3/2}$ as it introduces at most a $20\%$ error, within the uncertainty in $v_0$. The formula above only gives half the picture, as it does not account for the diffusion, or emptying, phase of the magnetic flux. Fortunately, in Section~\ref{sec:radvert_struc}, we found that the net outward velocity $v_\phi^{+}$ has the same form as the net inward velocity. Thus, we write
\begin{equation}
t_{\rm empty} \simeq \frac{2}{3}\frac{r_g}{v_0}\left(\frac{\delta\phi}{A}\right)^{2}=\frac{2}{3}\frac{r_g}{v_0}\left(\sigma_\phi\right)^{2}.
\end{equation}

The recurrence timescale of flux eruptions is then the sum of the emptying and refilling timescales:
\begin{align}
t_{\rm rec} &= t_{\rm empty}+t_{\rm refill}\\
t_{\rm rec} &= \frac{4}{3}\sigma_\phi^{2}\left(\frac{r_g}{v_0}\right)
\end{align}
which, for the values computed in this manuscript, yields
\begin{equation}
t_{\rm rec} \sim 800-2000\,\,r_g/c \quad \rm{for\,\, thin\,\,and\,\,thick\,\, disks}.
\end{equation}
These values are in full agreement with previous literature and with our own simulation results \citep{tchekhovskoy_efficient_2011,avara_efficiency_2016,scepi_magnetic_2023}. To our knowledge, this is the first analytical calculation to accurately predict this timescale. Finally, we note that flux eruptions are stochastic in nature; $t_{\rm rec}$ should be interpreted as an average recurrence timescale rather than a precise periodicity. We believe this stochasticity is naturally incorporated into our model through the choice of $\delta\phi = 2\sigma_\phi$; rarer events with different $\delta\phi$ values can be understood as statistical fluctuations in the amount of flux lost during each eruption.

\section{Turbulent structure in Magnetically arrested disks}
\label{sec:turbulent_struc}
As shown above, the cycles of diffusion and advection appear to be triggered or influenced by flux eruptions. To investigate this further, we analyze the structure of the turbulent magnetic diffusivity to determine whether its features resemble those of flux eruptions and whether it responds to such events. We first estimate, for the first time, the turbulent diffusivity in MADs to study its evolution during these episodes, which also enables us to measure the turbulent Prandtl number in MADs for the first time. Finally, in a more technical section, we examine the azimuthal structure of magnetic flux diffusion to identify possible non-axisymmetric features associated with flux eruptions.

\subsection{Effective resistivity}
\label{sec:eff_res}
\begin{figure*}
    \centering
    \includegraphics[width=0.9\textwidth]{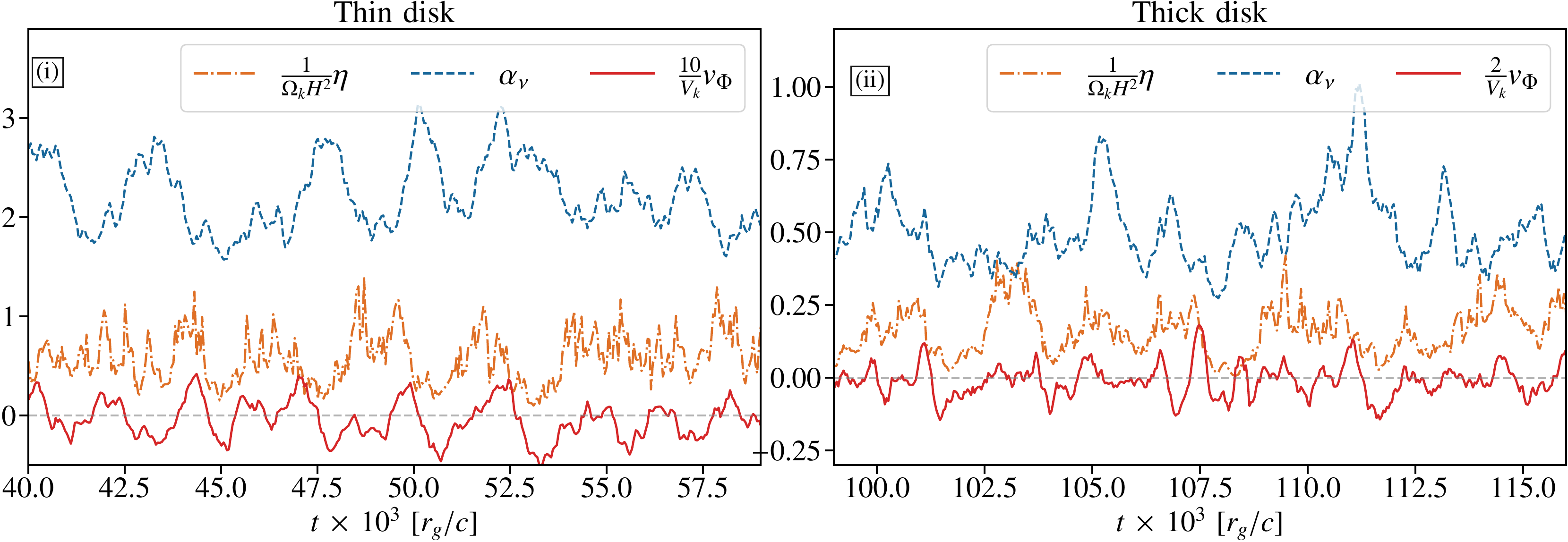}
    \caption{Temporal variability of the turbulent resistivity, $\eta$, averaged around the disk midplane at $r=10\,r_g$ and normalized to $\Omega_k H^2$, for the thin (i) and thick (ii) disk models. The quantity $\eta/(\Omega_k H^2)$ exhibits large oscillations in magnitude. Overplotted is the total magnetic flux transport velocity, $v_\Phi$, normalized to the local Keplerian velocity and multiplied by a factor of $10$ (thin disk) or $2$ (thick disk) for visibility. Maxima in $\eta$ correspond to episodes of magnetic flux diffusion, while minima coincide with flux advection, showing this correlation in both thin and thick disks. The Newtonian $\alpha_\nu$ parameter \citep{shakura_black_1973} is also shown, which displays strong variability but less than $\eta$. However, the variability of $\alpha_\nu$ is not as well correlated with flux eruptions as $\eta$.}
    \label{fig:eta_time}
\end{figure*}

\begin{figure}
    \centering
    \includegraphics[width=\columnwidth]{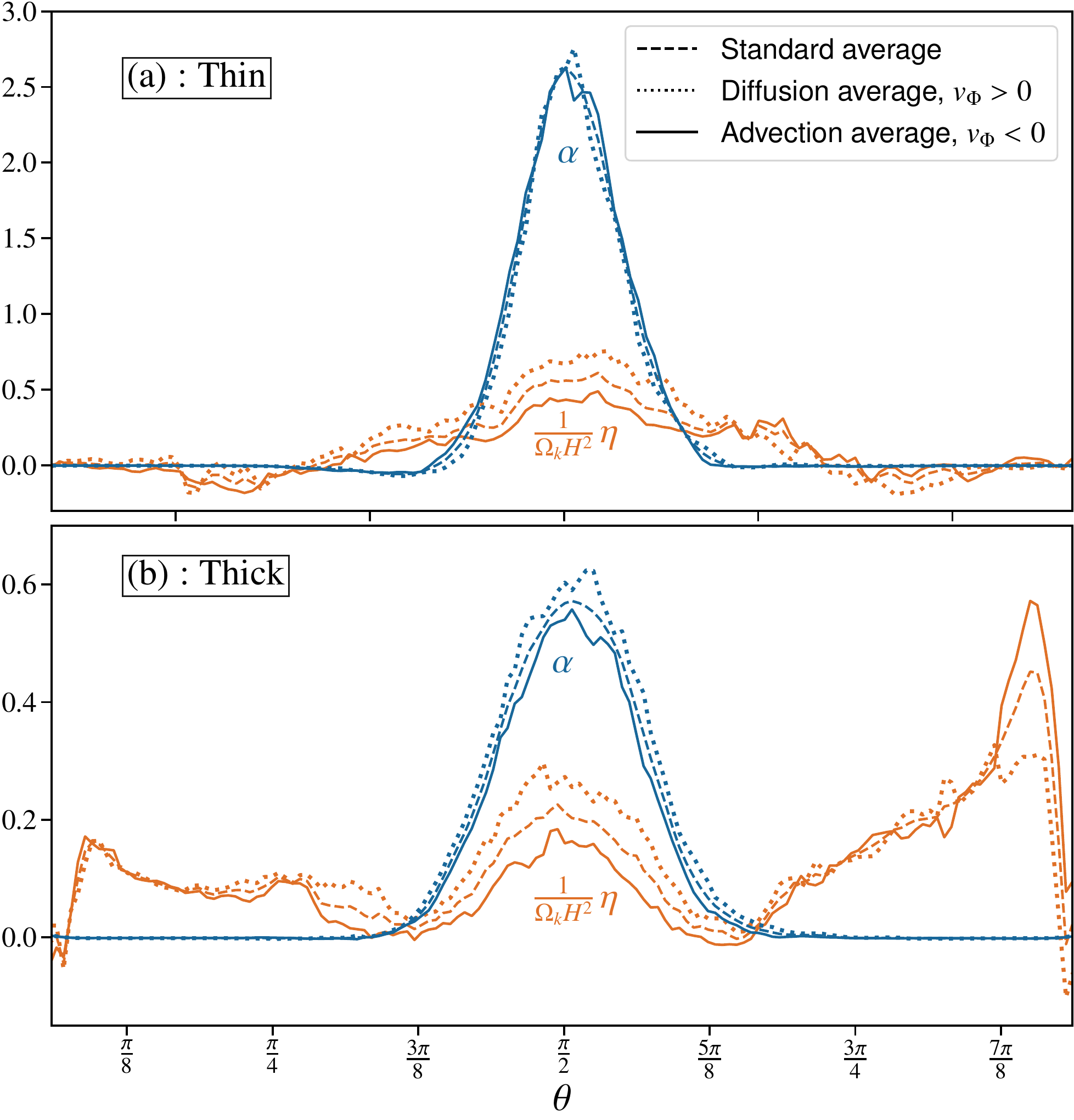}
    \caption{Time-averaged vertical profiles of $(1/\Omega_k^2H)\eta$ and $\alpha$ at $r=6\,r_g$ for thin (a) and thick (b) disks. Both show roughly Gaussian profiles, with resistivity broader than turbulent viscosity. The turbulent $\alpha$ is larger in magnitude, consistent with $\mathcal{P}_m>1$. During diffusion phases ($\eta^+$, $\alpha^+$), resistivity increases by about $50\%$, while during advection phases ($\eta^-$, $\alpha^-$) it decreases by a similar amount. Thus, diffusion phases are associated with enhanced resistivity, whereas advection phases coincide with reduced resistivity. The thick disk models show a modest change in turbulent viscosity, $\alpha$, with values enhanced by a factor of $1.1$ during diffusion events and reduced by a factor of $1.1$ during advection events. In contrast, we do not see significant changes in $\alpha$ between diffusion and advection phases for the thin disk model.}
    \label{fig:eta_vert}
\end{figure}

In this section we use a mean-field approach to study the diffusion of the large-scale magnetic field through the lens of a turbulent resistivity emerging from disk turbulence. For simplicity, we adopt a Newtonian framework\footnote{While it is, in principle, possible to measure mean-field resistivity in a relativistic setting, this requires ambiguous choices regarding the displacement current ($\partial_tE^\varphi$) and whether it should be treated as purely turbulent, purely laminar (i.e., large scale), or a combination of both.}. In what follows, we define Newtonian-like quantities using the standard hatted notation, $X\hatin{y} = X^y\sqrt{g_{yy}}$, where $X$ is a given quantity and $y$ is a spatial direction ($r$, $\theta$, $\varphi$). If $X$ is a velocity, we also divide by $u^t$.

We adopt the following simple closure for modeling the turbulent emf:
\begin{equation}
\mathcal{E}\hatin{\varphi} = -\eta \mean{J\hatin{\varphi}}.
\end{equation}
In full generality, one would expect $\mathcal{E}\hatin{\varphi}$ to depend on all components of the magnetic field $B\hatin{i}$ and their derivatives. As we will show, however, this closure provides a good approximation (see Appendix~\ref{A:fits}). Moreover, the presence of a strong large-scale magnetic field makes the inclusion of $\alpha_{\rm dynamo}$, shear-current, or turbulent pumping terms unnecessary, since a large-scale field is already present from the outset. Finally, the presence of a current sheet around the disk midplane favors terms that depend on $J\hatin{i}$.

To estimate the turbulent resistivity, we follow \cite{jacquemin-ide_magnetorotational_2024} with a simplified approach. Our procedure is as follows:
(1) We analyze the correlations between $\mathcal{E}\hatin{\varphi}$ and $J\hatin{\varphi}$ using
\begin{equation}
\label{eq:cp}
C_{\rm or}(X,Y) = \frac{\int\limits^{t_2}_{t_1} X(t')Y(t'),\mathrm{d}t'}{\sqrt{\int\limits^{t_2}_{t_1} X^2(t'),\mathrm{d}t' \int\limits^{t_2}_{t_1} Y^2(t'),\mathrm{d}t'}}.
\end{equation}
We compute this correlation for every $\theta$ at $r=6\,\,r_g$ (and confirm similar results at other radii) for both thin and thick disk simulations. We find maximum values of $C_{\rm or}(\mathcal{E}\hatin{\varphi},\mean{J\hatin{\varphi}})\simeq-0.45$ in both cases (see Appendix \ref{A:fits} and Fig.~\ref{fig:corel}). These maxima occur near the disk midplane, consistent with the vertical profiles shown in Section \ref{sec:radvert_struc}.

Our correlations $C_{\rm or}(\mathcal{E}\hatin{\varphi},\mean{J\hatin{\varphi}})$ are sufficiently strong to approximate the resistivity as
\begin{equation}
\eta \simeq C\,\frac{\int\limits^{t_2}_{t_1} \mathcal{E}\hatin{\varphi}(t')\mean{J\hatin{\varphi}}(t'),\mathrm{d}t'}{\int\limits^{t_2}_{t_1} \mean{J\hatin{\varphi}}^2(t'),\mathrm{d}t'},
\end{equation}
where $C$ is an order-unity coefficient used to maximize the accuracy of the resistivity estimate. Finally, we compare $\mathcal{E}\hatin{\varphi}$ with $\eta J\hatin{\varphi}$ and find that the latter provides a reliable approximation, as shown in Appendix \ref{A:fits}. Figure~\ref{fig:fits} confirms that our fits reproduce the emf signal with high accuracy.

In Figure \ref{fig:eta_time} we show the temporal variability of $\eta$, averaged around the disk midplane and evaluated at $r=6\,\,r_g$, normalized to $\Omega_k H^2$, for both thin (i) and thick (ii) disk models. The quantity $\eta/(\Omega_k H^2)$ is highly variable, undergoing large oscillations in magnitude, with maxima of about $0.2$ for the thick disk and $1.5$ for the thin disk, and minima of $\sim0.01$ and $\sim0.1$, respectively. Interestingly, the thin and thick disk models differ in magnitude by a factor of $5$–$10$. This discrepancy arises because the magnetic field, relative to the gas pressure, is overall stronger in the thin disk than in the thick disk. As a result, turbulence and resistivity scale with the field strength, producing a smaller signal in the thick case. Notably, this discrepancy does not appear in the diffusion velocity $v_{\rm diff}$, since that quantity is defined as a ratio that already incorporates the large-scale field.

To understand the role of the maxima and minima, we plot the total magnetic flux transport velocity $v_\Phi$, normalized to the local Keplerian velocity (and scaled by factors of $2$ (ii) or $10$ (i) for clarity), in Fig.~\ref{fig:eta_time} for thin (i) and thick (ii) disk models. We find a rough simultaneity between maxima of $\eta$ and episodes of magnetic flux diffusion, while minima of $\eta$ coincide with flux advection; this behavior is consistent across both thin and thick disks. Thus, the imbalance between advective and diffusive fluxes can be interpreted as dramatic changes in resistivity, probably driven or related to flux eruptions from the inner BH.

Finally, we also show the Newtonian \cite{shakura_black_1973} $\alpha_{\nu}$ parameter, defined as
\begin{equation}
\alpha_{\nu} = -\frac{1}{P_0}\mean{\delta B\hatin{r}\delta B\hatin{\varphi}},
\end{equation}
in Fig.~\ref{fig:eta_time}, where we ignore the Reynolds contribution since it is relatively unimportant.  We find that $\alpha_\nu$ also exhibits strong variability, though less pronounced than that of $\eta$, for both thin (i) and thick (ii) disk models. Unlike $\eta$, however, its variability is not as well correlated with flux eruptions, with correlations of only $\sim0.1$ for the thick disk and $\sim0.01$ for the thin disk, compared to $\sim0.45$ for $\eta$. 
Nonetheless, the presence of a correlation remains consistent with prior work \citep{chatterjee_flux_2022}.


From Fig.~\ref{fig:eta_time} we can estimate the turbulent Prandtl number associated with the turbulence driving magnetic flux transport. This quantity is an important property of the turbulence, as it has been shown to play a key role in large-scale magnetic field transport \citep{lubow_magnetic_1994}. We define the turbulent Prandtl number as
\begin{equation}
\mathcal{P}_m = \alpha \frac{H\Omega_k^2}{\eta},
\end{equation}
which is equivalent to the definitions commonly used in shearing-box studies.

For the thin disk model we find $\mathcal{P}_m\sim 1$ during diffusion phases and $\mathcal{P}_m\sim 5$ during advective phases. The thick disk model yields very similar values, with $\mathcal{P}_m\sim 1$ in diffusion phases and $\mathcal{P}_m\sim 7$ in advective phases. These results are consistent with shearing-box simulations of MRI turbulence, which find $\mathcal{P}_m\sim 1$–$3$ \citep{guan_turbulent_2009,lesur_turbulent_2009,fromang_turbulent_2009}, as well as with weakly magnetized global simulations, which give $\mathcal{P}_m\sim 3$–$5$ \citep{zhu_global_2018}. We therefore conclude that the turbulence properties constrained here fall within the expected range for MRI turbulence. That said, this does not exclude an additional role for flux eruptions in driving magnetic flux transport.

We now focus on the vertical structure of the resistivity for both thin and thick disk simulations. In Fig.~\ref{fig:eta_vert} we show the time-averaged profiles of $\tfrac{1}{\Omega_k^2H}\eta$ and $\alpha$, evaluated at $r=6\,\,r_g$ for thin (a) and thick (b) disks\footnote{We note that very similar structures are found at other radii.}. For both thin and thick disks, $\tfrac{1}{\Omega_k^2H}\eta$ and $\alpha$ exhibit roughly Gaussian-like vertical profiles. The thin disk shows a steeper profile for $\alpha$, as one might expect. The resistivity profiles, however, are broader than those of the turbulent viscosity. The magnitude of $\alpha$ is larger than that of $\tfrac{1}{\Omega_k^2H}\eta$, consistent with the results above showing $\mathcal{P}_m>1$ for both thin and thick disks.

To understand how the diffusion coefficients change between diffusion and advection events, we define the following time averages:
\begin{align}
X^+ &= \int\limits_{t_1}^{t_2}X\mathcal{H}(v_\Phi),\rm{d}t,\\
X^- &= \int\limits_{t_1}^{t_2}X\mathcal{H}(-v_\Phi),\rm{d}t,
\end{align}
where $\mathcal{H}$ is the Heaviside function. Here, $\eta^{+}$ corresponds to the resistivity during diffusion events, when the flux is transported outwards, and $\eta^{-}$ corresponds to the resistivity during advection events, when the magnetic field is advected inwards.

In Fig.~\ref{fig:eta_vert} we also show the vertical profiles of $\eta^{+,-}$ and $\alpha^{+,-}$ to examine how they vary during advection and diffusion phases. For both thin and thick disk models, we find that the resistivity increases by roughly $50\%$ during diffusion phases compared to the time-averaged value, while it decreases by about $50\%$ during advection phases. Thus, diffusion phases are associated with enhanced resistivity, whereas advection phases coincide with reduced resistivity. It remains unclear how this increase in turbulent resistivity is connected to flux eruptions themselves, but we speculate that flux eruptions may contribute to it by driving additional instabilities, such as interchange-like instabilities \citep{lubow_magnetic_1995}.

The thick disk models show a modest change in turbulent viscosity, $\alpha$, with values enhanced by a factor of $\sim1.1$ during diffusion events and reduced by a factor of $\sim 1.1$ during advection events. This is consistent with the results of \cite{chatterjee_flux_2022}, who also found enhanced angular momentum transport during flux eruptions, although our results suggest this effect is small magnitude. In contrast, we do not see significant changes in $\alpha$ between diffusion and advection phases for the thin disk model.

\subsection{Flux transport criteria}

The seminal work of \cite{lubow_magnetic_1994} introduced a simple analytic criterion for determining whether the magnetic flux is transported inward or outward. Thanks to the occurrence of both flux advection and diffusion events in MADs, these systems provide a valuable testing ground for assessing the accuracy of different analytical flux transport models. In this section, we compare various flux transport criteria to identify which best reproduces the behavior observed in MAD simulations.

We begin with the most accurate criterion and then outline the series of approximations required to derive the other criteria, ultimately arriving at the one proposed by \cite{lubow_magnetic_1994}. We also include a discussion of the recent criterion introduced by \cite{begelman_simple_2024}.

The most direct and reliable diagnostic of whether a system is advecting or diffusing magnetic flux is the ratio of the advection to diffusion velocities. Accordingly, we define
\begin{equation}
\label{eq:true_crit}
\mathcal{D} = \frac{v_{\rm diff}}{-v_{\rm adv}} = \frac{\thmean{\mathcal{E^\varphi}}}{-\thmean{\mathcal{A}}},
\end{equation}
against which all other flux transport criteria will be compared. When $\mathcal{D} > 1$, the large-scale field is being diffused; when $\mathcal{D} < 1$, it is being advected inward.

The next most reliable criterion uses a mean-field and Newtonian approximation to estimate the turbulent electromotive force, setting $\sqrt{g_{rr}g_{\theta\theta}}\thmean{\mathcal{E^{\varphi}}} = \thmean{\eta J\hatin{\varphi}}$, and defining
\begin{equation}
\label{eq:deta_crit}
\mathcal{D}_\eta = \frac{\thmean{\eta J\hatin{\varphi}}}{\thmean{\mathcal{A\hatin{\varphi}}}},
\end{equation}
where we define the Newtonian field advection term as $\mathcal{A\hatin{\varphi}} = \sqrt{g_{rr}g_{\theta\theta}}A^{\varphi}$.

\begin{figure*}
    \centering
    \includegraphics[width=0.9\textwidth]{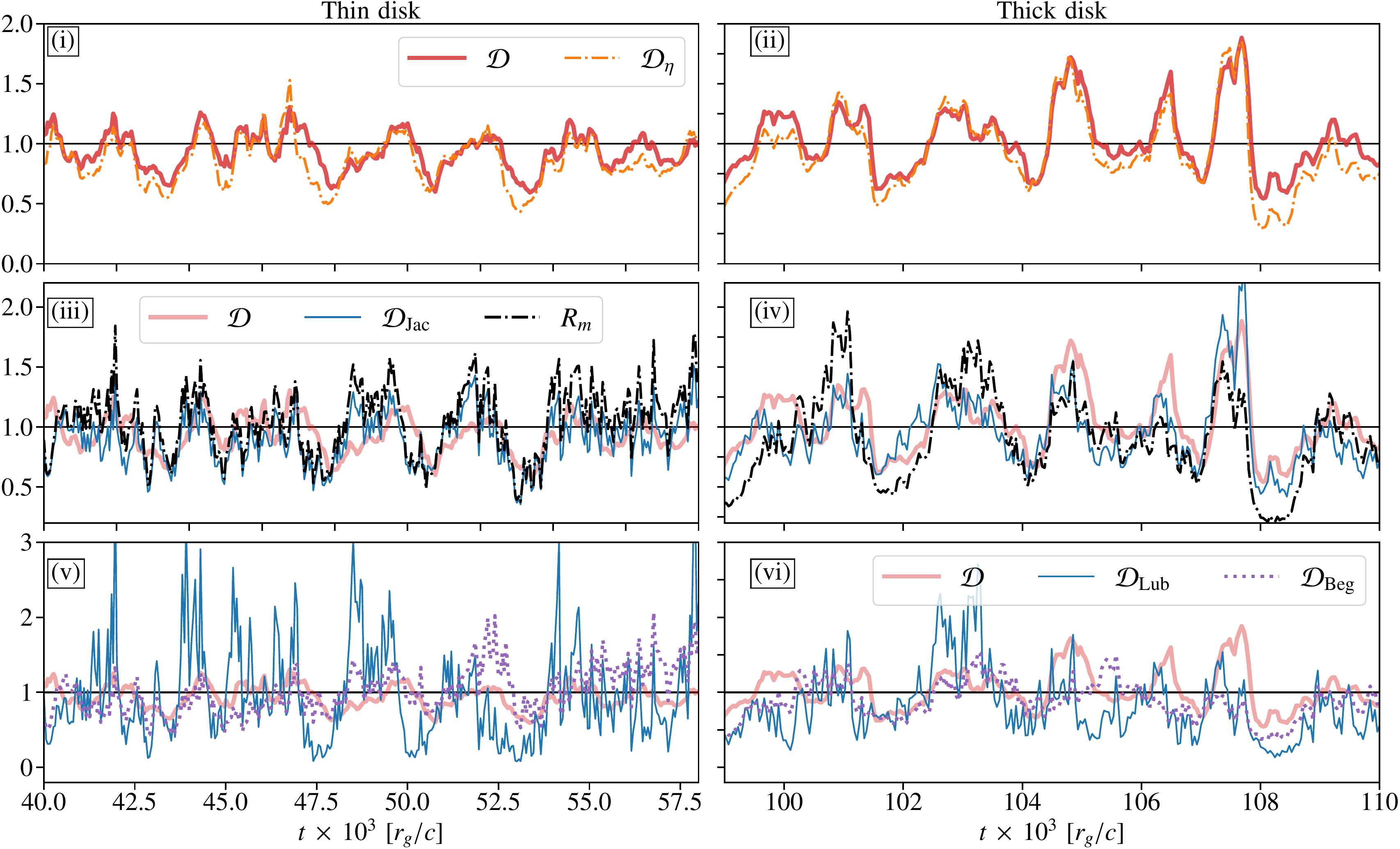}
    \caption{Comparison of different criteria for determining whether the system is advecting or diffusing large-scale magnetic flux. The true criterion, $\mathcal{D}$, is shown alongside several simplified criteria to assess their accuracy. Panels (i) and (ii) compare $\mathcal{D}$ with $\mathcal{D}_\eta$ at $r = 6\,r_g$ for the thin and thick disk models, respectively, showing striking agreement. Panels (iii) and (iv) compare $\mathcal{R}_m$ \citep{lubow_magnetic_1994} and the criterion $\mathcal{D}_{\rm Jac}$ derived here with $\mathcal{D}$; both track its evolution relatively well, particularly in the thick disk case, although $\mathcal{R}_m$ required an adjustment coefficient. Panels (v) and (vi) compare $\mathcal{D}_{\rm Lub}$ \citep{lubow_magnetic_1994} and $\mathcal{D}_{\rm Beg}$ \citep{begelman_simple_2024} with $\mathcal{D}$; $\mathcal{D}_{\rm Lub}$ is noisy, tends to overestimate both regimes, and shows sharp transitions, requiring only an order-unity coefficient, while the \cite{begelman_simple_2024} model stays closer to $\mathcal{D}$ in both phases but required a larger adjustment coefficient.}
    \label{fig:criteria}
\end{figure*}

We show both criteria, $\mathcal{D}$ and $\mathcal{D}_\eta$, evaluated at $r = 6\,r_g$ as functions of time in Fig.~\ref{fig:criteria}, for the thin (i) and thick (ii) disk models. The agreement between the two is striking.  This confirms that our mean-field model accurately describes magnetic field transport.

Further approximations can be applied to Eq.~\ref{eq:deta_crit} by expressing $J\hatin{\varphi} \simeq \frac{B\hatin{r}^{s}}{H}$, where $B\hatin{r}^{s} = \frac{1}{2}(|B\hatin{r}(\theta = \theta_1)| + |B\hatin{r}(\theta = \theta_2)|)$ is the large-scale radial field at the surface of the accretion disk\footnote{where $\theta_1$ and $\theta_2$ are chosen to be a few scale heights above and below the disk.}. We approximate
\begin{equation}
\thmean{\mathcal{A\hatin{\varphi}}} \simeq -\thmean{u\hatin{r}B\hatin{\theta}},
\end{equation}
neglecting the term associated with vertical displacements of the radial magnetic field, i.e., $\thmean{u\hatin{\theta}B\hatin{r}}$. These approximations yield the following criterion:
\begin{equation}
\label{Eq:deta_jac}
\mathcal{D}_{\rm Jac} = \frac{\eta B\hatin{r}^{s}}{H\thmean{u\hatin{r}B\hatin{\theta}}}.
\end{equation}

Assuming $B\hatin{r}^{s} \simeq B\hatin{\theta}$ allows us to recover the criterion from \cite{lubow_magnetic_1994}:
\begin{equation}
\label{eq:Rmlub}
\mathcal{R}_m = C_{\rm RM}\frac{\thmean{\eta}}{H|\thmean{u\hatin{r}}|},
\end{equation}
where we denote this coefficient $\mathcal{R}_m$ due to its similarity with the magnetic Reynolds number. Note, however, that this is an effective Reynolds number relying on turbulent, rather than microphysical, resistivity. We introduce the constant coefficient $C_{\rm RM}$ to account for the limitations of this criterion—without it, the expression would not closely match $\mathcal{D}$. We find that setting $C_{\rm RM} \simeq 7$ yields good agreement with the more accurate diagnostics. 

In Figure~\ref{fig:criteria} (iii and iv), we compare the criteria $\mathcal{R}_m$ and $\mathcal{D}_{\rm Jac}$ with the true criterion $\mathcal{D}$ for thin (iii) and thick (iv) disk models. We find that both simplified models follow the evolution of the true criterion relatively well, especially in the thick disk case. For the thin disk, however, both models tend to overestimate the duration of diffusion phases (i.e., when $\mathcal{D} > 1$). We observe that $\mathcal{D}_{\rm Jac} \simeq \mathcal{R}_m$, particularly for the thin disk model. However, it should be noted that in order for $\mathcal{R}_m$ to match the data, we had to introduce a correction factor of $C_{\rm RM} \simeq 7$. As expected, the criterion with fewer approximations better reproduces the numerical behavior.

\cite{lubow_magnetic_1994} introduced a more restricted criterion applicable when accretion is primarily driven by turbulent $\alpha$-stresses. In that case, one can assume $u\hatin{r} \sim \alpha H \Omega_K \frac{H}{R}$ to derive the following criterion:
\begin{equation}
\mathcal{D}_{\rm Lub} = C_{\rm Lub} \frac{R}{H} \frac{\thmean{\eta}}{\thmean{\alpha} H^2 \Omega_K},
\end{equation}
where, again, we include the constant coefficient $C_{\rm Lub}$ to compensate for the limitations of the model. We find that this criterion best matches the numerical data when $C_{\rm Lub} \sim 0.5$.

Finally, \cite{begelman_simple_2024} proposed a different type of criterion based on their slab model for advection. Instead of directly comparing advection and diffusion of the large-scale field, their criterion depends on how both processes scale with the field strength. It is given by
\begin{equation}
\mathcal{D}_{\rm Beg} = C_{\rm Beg} \frac{\eta}{H V^{A}\hatin{\theta}},
\end{equation}
where we define
\begin{equation}
V^{A}\hatin{\theta} = \frac{|\thmean{B_\theta}|}{\sqrt{\thmean{\rho}}}.
\end{equation}
As before, we include the constant coefficient $C_{\rm Beg}$ to adjust for the model's simplifications. We find that the best agreement with the data occurs for $C_{\rm Beg} \simeq 40$.

In Figure~\ref{fig:criteria} (v and vi), we compare the criteria $\mathcal{D}_{\rm Lub}$ and $\mathcal{D}_{\rm Beg}$ to the true criterion $\mathcal{D}$ for thin (v) and thick (vi) disk models. We find that $\mathcal{D}_{\rm Lub}$ is very noisy and tends to overestimate both diffusion and advection phases, with very sharp transitions between the two regimes. However, despite not accurately reproducing the evolution of the curves, it only required a coefficient of order unity. We attribute the inaccuracies of this model to its assumption that all accretion arises from viscous torques, completely ignoring the important role of large-scale torques in MADs \citep{scepi_magnetic_2023,manikantan_winds_2023}.

The \cite{begelman_simple_2024} model performs better, staying closer to $\mathcal{D}$ during both advection and diffusion phases. This is commendable, considering that the criterion contains no explicit information about advection and models it only through the magnitude of the Alfvén velocity, $V_a$. Nonetheless, the model required a large coefficient to match the data, $C_{\rm Beg} \simeq 40$, which might stem from limitations of the slab model for disks with $h/R \geq 0.1$; it could also be a consequence of the simplified assumptions used for turbulent diffusion.

\subsection{Azimuthal structure}
\label{sec:azimuth_struct}

We now investigate the azimuthal structure of magnetic flux diffusion. Historically, flux eruptions have been recognized as strong non-axisymmetric features that redistribute magnetic flux. However, whether they contribute to the transport of large-scale, axisymmetric, magnetic fields has remained unclear, as has the mechanism by which such transport might occur. In this section, we explore the connection between non-axisymmetric structures and magnetic field transport.
This section presents a technical analysis and can be skipped by readers interested only in the main results.

\begin{figure}
    \centering
    \includegraphics[width=\columnwidth]{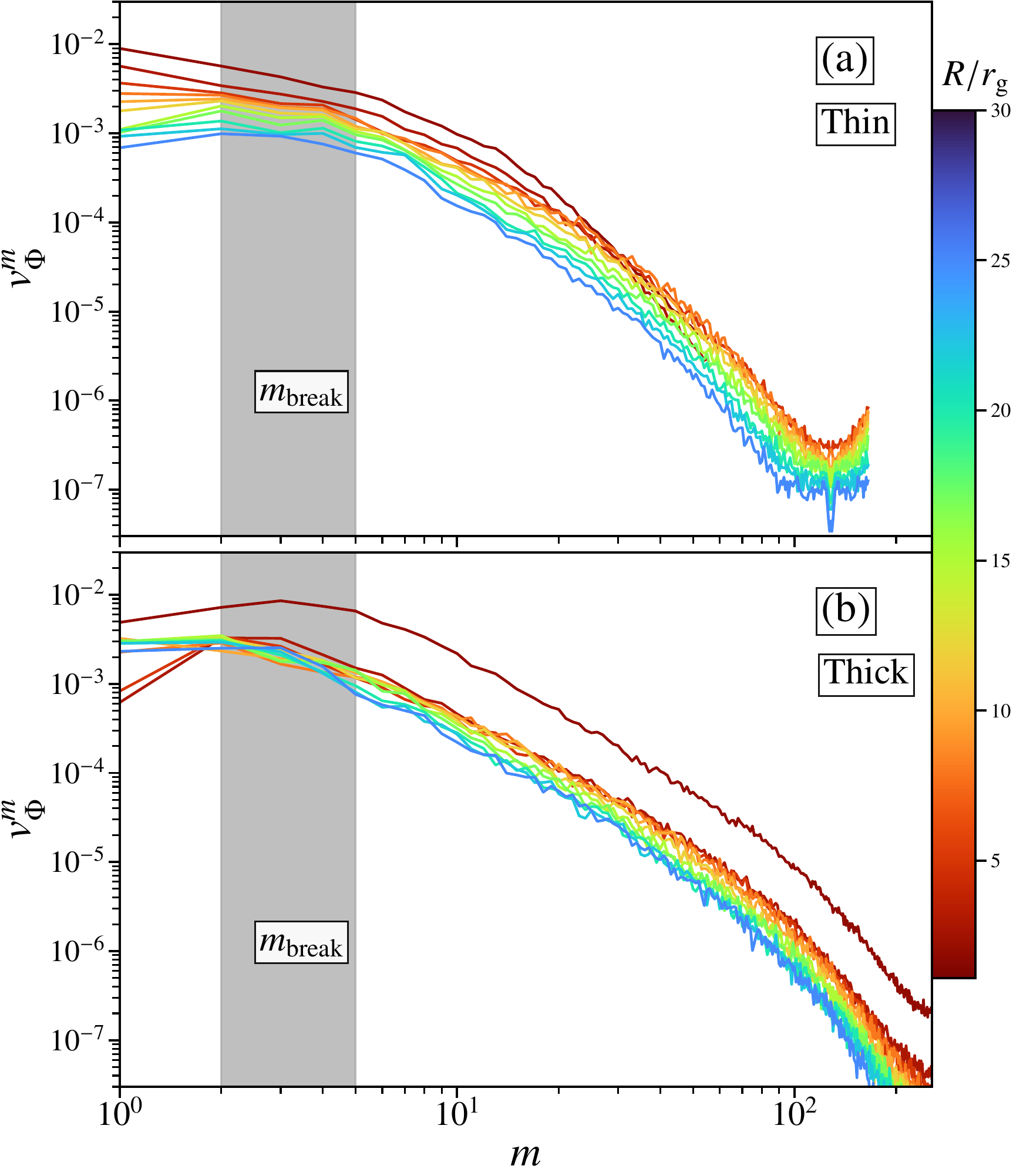}
    \caption{Distribution of $v_\Phi^m$ versus azimuthal wavenumber $m$ at several radii in the thin (a) and thick (b) disk models. In both cases, the $m$-dependence is largely independent of radius, with all curves showing a power-law trend from $m \sim 5$ to $m \sim 100$ and a break near $m_{\rm break} \sim 3$ indicated by the gray region. Some radii show a pronounced peak at $m_{\rm break}$, while others peak at $m \lesssim m_{\rm break}$. Any azimuthal mode number $m$ equal to or greater than the Nyquist limit, $m_{\rm nyn} \sim 0.5 N_\varphi$, is unreliable and should be excluded from the analysis.}
    \label{fig:pow_struc}
\end{figure}
To do this we use a Fourier decomposition of the different modes contributing to the advection and diffusion of the large scale axisymmetric field , following \cite{jacquemin-ide_magnetorotational_2024}. Using Parseval's theorem we write
\begin{equation}
    E^\varphi = \sum\limits_{m=0}^\infty E^{\varphi,(m)},
    \label{eq:parseval}
\end{equation}
where
\begin{equation}
   E^{\varphi,(m)} = \mathcal{R}\left[u^{r,m}\,\, b^{\theta,m}|^{*}-u^{\theta,m}b^{r,m}|^{*}\right],
    \label{eq:EMF_mode}
\end{equation}
and $\mathcal{R}$ represents taking the real part, $X|^{*}$ is the complex conjugate of $X$ and ${u}^{i,m}$ and ${b}^{i,m}$ are the spectral coefficients of the $m$-mode in the Fourier expansion of ${u^i}$ and ${b^i}$, 
\begin{align}
\label{Eq:upFour}
    {u^{i}} &= \sum\limits_{m=0}^\infty{u}^{i,m}e^{im\varphi}\\
    {b^i} &= \sum\limits_{m=0}^\infty{b}^{i,m}e^{im\varphi}.
    \label{Eq:BpFour}
\end{align}
Note that $\mathcal{E}_\varphi^{(m)}$ represent the axisymmetric effect of nonlinear interaction of non-axisymmetric $m$-modes\footnote{{In practice we never compute the Fourier modes of the EMF. We only extract the Fourier modes of the velocity and magnetic fields  to calculate $\mathcal{E}_\varphi^{(m)}$ in Eq.~(\ref{eq:EMF_mode})}.}.{ The $^{(m)}$ superscript denotes nonlinear-correlations on the mean ($\varphi$-averaged) EMF due to non-zero $m$ fluctuations of ${u^i}$ and ${b^i}$, see more details in \cite{jacquemin-ide_magnetorotational_2024}.

Finally, we define the $m$-modes of the flux transport velocity
\begin{align}
    v_\Phi &=\frac{1}{\mean{B^\theta}} \sum\limits_{m=0}^\infty E^{\varphi,(m)},\\
    v_\Phi^{(m)} &= \frac{1}{\mean{B^\theta}}E^{\varphi,(m)}. 
\end{align}
From the expressions above we can deduce that 
\begin{align}
     v_{\rm adv} &= v_\Phi^{(m=0)},\\
    v_{\rm diff} &= \sum\limits_{m=1}^\infty v_\Phi^{(m)};
\end{align}
therefore, in this section we only focus on modes $m\geq1$, as $m=0$ was well explored in the sections above.

Any azimuthal mode number $m$ equal to or greater than the Nyquist limit, $m_{\rm nyn}\sim 0.5 N_\varphi$, where $N_\varphi$ is the azimuthal grid size, is unreliable and is excluded from the analysis. According to the Nyquist–Shannon sampling theorem, modes above this threshold suffer from aliasing and do not represent physical structures.
For safety sake we do not analyze modes $m\geq0.8m_{\rm nyn}$. Finally, we note that $m_{\rm nyn}$ differs between thin and thick disk models and, in the thin disk case, varies with radius due to AMR de-refinement at small radii ($r \leq 5\,r_g$), which lowers the effective resolution of that simulation in the inner regions. For this reason the trends in $m$ for radii $r\leq5\,\,r_g$ in the thin disk model are to be taken with a grain of salt. 

In Figure \ref{fig:pow_struc}, we show the distribution of  $v_\Phi^m$, as a function of $m$, plotted for several radii in both the thin (a) and thick (b) disk models and normalized to $V_k$. We observe that for both models, the azimuthal wave number dependence appears largely independent of radius, with different radii producing very similar azimuthal $m$-mode dependency. All curves exhibit a consistent structure: a clear power-law trend between $m \sim 5$ and $m \sim 100$, with a negative power-law exponent of roughly $2-3$. Additionally, all curves show a noticeable break near $m_{\rm break} \sim 3$, with some displaying a distinct peak at this value and others peaking at $m \lesssim m_{\rm break}$. This power-law break, $m_{\rm break}$, is indicated by the gray region in Fig.~\ref{fig:pow_struc}.

We are tempted to associate the power-law break at $m_{\rm break} \simeq 2$–$5$ with the large-scale flux eruption features commonly seen in global GRMHD simulations of MADs. These eruptions are typically non-axisymmetric and dominate at large azimuthal spatial scales, corresponding to low azimuthal mode numbers around $m \sim 4$. Thus, the break observed at $m \sim 3$ could reflect the scale at which energy is injected into the system by these eruptions. This energy may then cascade to smaller scales, contributing to the diffusion of large-scale magnetic flux. However, even though $v_\Phi^m$ peaks near $m \sim 3$, this contribution is small compared to the combined effect of the many higher-$m$ modes. This is evident from the fact that $v_\Phi^{(m=2,3,4,5)} \ll v_{\rm diff} = \sum_{10}^{\infty} v_\Phi^m$, indicating that the numerous smaller-scale modes dominate the overall flux diffusion.

We stress that flux eruptions are not the sole mechanism by which magnetic flux is transported to larger radii. Even in the absence of visible eruptions, $v_{\rm diff}$ remains significant—indicating that diffusion is continuously active. 
However, eruptions appear to trigger a transition to outward flux diffusion, likely because the system is already near equilibrium and the eruption pushes it beyond this point. Our interpretation of this is that flux eruptions are coupled to the disk turbulence, which is likely driven MRI, and therefore enhance dissipation enabling flux diffusion. As we showed in the previous section, the measured turbulent Prandtl number, $\mathcal{P}_m$, closely matches the one measured in MRI-driven turbulence.

\section{Conclusion and Discussion}
\label{sec:conc_disc}
\subsection{Summary}

Magnetically arrested disks have emerged as a popular disk model due to their ability to reproduce some of the observational features of low-luminosity AGN. In particular, the outbursts of AGN like Sgr A* and $\gamma$-ray flares from other AGN can be explained by features of MADs \citep{dexter_sgr_2020,hakobyan_radiative_2023}. Indeed, MADs undergo powerful flux eruptions in which magnetic fields are ejected from the event horizon, potentially serving as drivers of the luminosity changes observed in nature. Our objective was to understand what drives flux eruptions and how they serve as a form of communication between the BH and the accretion disk. We focus on both thin and thick accretion disk, but find that for the disk thicknesses explored, the story of flux eruptions, advection, and diffusion of the large-scale field appears mostly unchanged.

In this work, we introduce a new formalism for magnetic field transport in 3D GRMHD accretion disks around BHs. The formalism is based on an advection equation, but with a net flux transport velocity, $v_\Phi = v_{\rm adv} + v_{\rm diff}$, that depends on both advective ($v_{\rm adv}$) and diffusive ($v_{\rm diff}$) processes within the accretion disk. We have shown that this velocity accurately reproduces the behavior of the large-scale axisymmetrized field, with its sign indicating whether the large-scale magnetic flux is in the process of being advected toward the inner radii or diffusing outward to larger radial scales (see Fig.~\ref{fig:model_valid}). 

We find that the complete dynamical system is in a statistical quasi-steady state, in the sense that advective and diffusive processes are roughly balanced, and both are larger than the net flux transport velocity; i.e., $|v_{\rm adv}| \simeq |v_{\rm diff}| \gg |v_\Phi|$. Episodes of flux advection and diffusion can be interpreted as slight imbalances within this equilibrium (see Figs.~\ref{fig:vdiff_vadv},~\ref{fig:rad_struc_v},~\ref{fig:vert_struc_v} ). We find that, due to this balance, the net inward $v_\Phi^-$ and outward $v_\Phi^+$ flux transport velocities are an order of magnitude smaller than the advective $v_{\rm adv}$ and diffusive $v_{\rm diff}$ velocities. This insight is summarized in Fig.~\ref{fig:diagram}.
We find that flux eruptions are correlated with flux diffusion phases (see Fig.~\ref{fig:r_t_erupt}), and likely enhance turbulent diffusivity within the accretion disk, which then drives magnetic field diffusion (see more below). 

We then use this new insight to analytically compute the recurrence timescale of flux eruptions using the net inward ($v_\phi^-$) and outward ($v_\phi^+$) velocities, their radial dependence, and also the radial dependence of the large-scale magnetic flux anchored on the accretion disk, $\phi$ (see Fig.~\ref{fig:fit_phi}).  We find that the recurrence timescale is given by
\begin{equation}
t_{\rm rec} = \frac{4}{3}\left(\sigma_\phi\right)^{2}\left(\frac{r_g}{v_0}\right)\simeq 1500\,\,r_g/c,
\end{equation}
where $\sigma_\phi$ is the standard deviation of the large-scale flux around the event horizon, representing how much flux is lost during an eruption.
The fact that this formula reproduces the correct average recurrence timescale for flux eruptions lends further credibility to the formalism developed in this paper. To our knowledge this is the first time this timescale has been computed analytically.

We then focus on better understanding the turbulence that drives the diffusion of the large-scale field during diffusion episodes. To do this, we compute the effective resistivity produced by the turbulence acting on the large-scale magnetic field. This is the first time turbulent resistivity has been measured in MADs. We find that the effective magnetic Prandtl number, the ratio of turbulent viscosity to turbulent resistivity, is consistent with MRI-driven turbulence , i.e., $\mathcal{P}_m \simeq 1 - 5$ (see Fig.~\ref{fig:eta_time},~\ref{fig:eta_vert}), as measured in the MRI shearing box literature \citep{guan_turbulent_2009,lesur_turbulent_2009,fromang_turbulent_2009}.

At the same time, we do not rule out the role of flux eruptions in shaping the turbulence and contributing to the diffusion of the large-scale field. In fact, we find that turbulent resistivity increases during flux eruption events (see Fig.~\ref{fig:eta_time},~\ref{fig:eta_vert}). When examining the non-axisymmetric modes that contribute to magnetic field diffusion, we observe a clear spike around $m \simeq 3 - 5$, which corresponds to the expected scales of flux eruptions. However, even though these modes show enhanced activity, their overall contribution is still outweighed by the more numerous smaller-scale modes. These smaller-scale modes remain the dominant driver of large-scale magnetic field diffusion. We believe flux eruption might be enhancing the local turbulent dissipation, leading to large scale diffusion of the magnetic field.

Finally, we test several analytic criteria intended to determine whether the magnetic field is being advected inward or diffused outward. We find that the classical effective Reynolds number criterion proposed by \cite{lubow_magnetic_1994} approximately reproduces the data, although it requires scaling by a large constant. Their second criterion, based on the turbulent Prandtl number, performs without a large scaling factor but produces a noisier signal that does not follow the data as closely. The more recent criterion proposed by \cite{begelman_simple_2024} also gives a reasonable match to the data, again with the help of a large constant coefficient. Motivated by these results, we propose a new criterion:
\begin{equation}
\label{Eq:deta_jac_conc}
\mathcal{D}_{\rm Jac} = \frac{\eta B\hatin{r}^{s}}{H\thmean{u\hatin{r}B\hatin{\theta}}},
\end{equation}
which we find tracks the simulation data with significantly improved accuracy.

\subsection{Discussion}

\emph{What drives the turbulence?} 
It has been argued that MRI may not be the main driver of turbulence in MADs \citep{white_resolution_2019}. The reasoning is that as the magnetic field strengthens, field lines become increasingly difficult to bend on small scales. Beyond a certain strength, the smallest possible bend exceeds the disk scale height, fully stabilizing the MRI \citep{balbus_instability_1998}. 
The problem with this argument is that it oversimplifies the complexities of simulated accretion disks by relying on linear analyses of axisymmetric MRI: (1) It neglects non-axisymmetric MRI modes\footnote{Sometimes called super Alfv\'enic rotational instability \citep[SARI:][]{goedbloed_super-alfvenic_2022}.} which are present in accretion disks and remain unstable at much higher magnetic field strengths \citep{ogilvie_non-axisymmetric_1996,das_instability_2018,begelman_what_2022,goedbloed_super-alfvenic_2022,brughmans_parametric_2024,brughmans_visual_2025}; (2) it overlooks that linear axisymmetric MRI in stratified disks can also persist at higher field strengths \citep{latter_mri_2010}; 

If MRI is suppressed, other instabilities—such as interchange-like instabilities \citep{lubow_magnetic_1995} or tearing-like instabilities \citep{begelman_saturation_2023}—could take over angular momentum transport. However, not all instability-driven turbulence can transport angular momentum; effective transport requires coherent motions in both the poloidal and toroidal directions, generating strong Reynolds and Maxwell stresses. Whether interchange-like or tearing-like instabilities can fulfill this role remains unproven. Furthermore,  both instabilities feed on density, pressure, or magnetic field gradients, which generally provide a smaller energy reservoir than shear in most regions of the disk—except very close to the event horizon.

In this work, we find that turbulence in MADs exhibits a magnetic Prandtl number similar to that of MRI turbulence, $\mathcal{P}_m \sim 3$ \citep{guan_turbulent_2009,lesur_turbulent_2009,fromang_turbulent_2009}. \cite{begelman_what_2022} and \cite{scepi_magnetic_2023} further showed that the product of $\alpha$ and $\beta$ in MADs is also consistent with MRI turbulence. Overall, our results suggest that MRI—or a related instability, such as its non-axisymmetric version \citep{brughmans_visual_2025}—plays a key role in driving turbulence in MADs. Nonetheless, interchange-like instabilities are likely still present but may play a supporting role, as they likely occur during flux eruptions that modulate large-scale transport of the magnetic field and, consequently, turbulent diffusion in MADs. Indeed, interchange-like instabilities might regulate field transport instead of angular momentum transport.

\emph{Consequences of flux transport in observations of jetted accretion disks:}
Understanding the efficiency of flux transport is crucial for identifying the physical conditions required for jet launching. If magnetic flux transport is efficient, it poses an observational puzzle: not all accreting objects produce jets—only about $10\%$ of AGN are jetted \citep{padovani_microjansky_2011,zamaninasab_dynamically_2014}, XRBs launch jets only during specific outburst phases \citep{done_modelling_2007}, and only a few TDEs produce prompt jets \citep{burrows_relativistic_2011}. Below, we discuss the limits and assumptions of our framework that lead to the prediction of more efficiently magnetized—and therefore more frequently jetted—BHs than current observations suggest.

In this work we find that inward flux transport may be easier to achieve than previously thought. \cite{lubow_magnetic_1994} argued that flux transport could be impossible for most disks thicknesses, formulating the criterion $D_{\rm Lub}<1$, where $D_{\rm Lub}=\frac{R}{H}\frac{1}{\mathcal{P}_m}$. Since $\mathcal{P}_m \sim 3$, this condition is very difficult to satisfy in most disks, $h/R\leq0.3$. In this work, we propose a more accurate criterion for field advection: $D_{\rm Jac}<1$, where $D_{\rm Jac}=\frac{\eta}{Hu_r}\frac{B_r^{s}}{B_\theta}$. This condition is easier to satisfy because $u_r$ includes accretion driven not only by turbulent torques but also by large-scale magnetic torques \citep{ferreira_magnetized_1995,scepi_magnetic_2023}, which produce higher inflow velocities than turbulence alone. We show that MADs with $h/R\geq0.1$ can advect magnetic flux. Even thinner MADs with $h/R\simeq0.03$ have also been observed, implying that some level of flux transport must occur at those thicknesses \citep{liska_formation_2022,scepi_magnetic_2023}.
We also find that flux transport is concentrated near the disk midplane, indicating that coronal accretion—previously hypothesized as a mechanism for flux transport in accretion disks \citep{rothstein_advection_2008}—is not required in MADs. Nonetheless, coronal advection has been shown to play an important role in more weakly magnetized disk models \citep{beckwith_transport_2009, jacquemin-ide_magnetic_2021}, where a magnetically dominated corona forms above the disk. Such configurations have also been observed in radiatively cooled thin disks \citep{liska_magnetic_2024, zhang_radiation_2025}.

One possible explanation for the efficient magnetic field transport—and the resulting higher occurrence of jetted accretion disks than observations suggest—is that strong magnetic fields naturally promote jet launching through efficient field transport, while non-jetted systems correspond to more weakly magnetized disks. \cite{jacquemin-ide_magnetic_2021} found that the flux transport velocity depends strongly on the initial disk magnetization, becoming very small for weakly magnetized disks ($v_\phi\sim10^{-4}V_k$), and also varies significantly with disk thickness, as predicted by \cite{lubow_magnetic_1994}. Furthermore, \cite{mishra_strongly_2020} found no measurable flux transport in thin, $h/R=0.05$, non-MAD disks, consistent with the trends identified by \cite{jacquemin-ide_magnetic_2021}. Therefore, MADs with $h/R\geq0.1$ may represent the ideal conditions for efficient flux transport and, consequently, jet production.

Another possible explanation for jet shutdown during certain phases of accretion disk evolution is the cancellation of opposing polarities of the large-scale vertical magnetic field \citep{parfrey_black_2015, scepi_magnetic_2021}. This process could cause accretion disks to spend significant periods without net magnetic flux.

Even with these limitations, flux transport remains a very promising mechanism to explain changes in jet efficiency during X-ray binary state transitions \citep{ferreira_unified_2006,marcel_unified_2019,liska_formation_2022,liska_magnetic_2024,naethe_motta_black_2025}; see however \cite{scepi_thermal_2024}.

\emph{MADs as the final state of a disk with a large flux reservoir.} The fact that the MAD state appear as the limit where the flux transport velocity $|v_\Phi|\sim0$ suggests they can be understood as the final quasi–steady state of an accretion disk with a large magnetic flux reservoir. When such a disk forms, it first goes through an advection phase with $|v_\Phi|\simeq |v_{\rm adv}|\ll |v_{\rm diff}|$. As flux saturates on the black hole, the magnetic field strength in the disk increases, boosting turbulent diffusivity and driving the system toward $|v_{\rm adv}|\simeq |v_{\rm diff}|\gg |v_\Phi|$. This picture is consistent with our results and those of \cite{jacquemin-ide_magnetic_2021}, where $|v_\Phi|$ approaches zero as the system transitions into a MAD state, or Newtonian MAD for \cite{jacquemin-ide_magnetic_2021}. The MAD could then be understood as a kind of dynamical attractor, with the amount of large-scale magnetic flux acting as the control parameter that determines whether or not the accretion disk becomes MAD.

\emph{Observational consequences.} Our work provides a solid theoretical framework for understanding and modeling flux eruptions in MADs. This framework could be applied to model AGN variability using simplified, semi-analytical approaches that can be more directly compared to observations, while also yielding information on the properties of accretion disks—for example, statistics on advection and diffusion rates across different disk outbursts.

Within this study, we have extended our understanding of magnetic field transport and its role in the formation and steady state of MADs. These results also offer insight into the variability and outbursts of AGN disks and point to a promising avenue for future observations from transient observatories, such as the Vera Rubin Observatory.

\section*{acknowledgments}
JJ acknowledges insightful conversations with Nick Kaaz, Danat Issa and Prasun Dhang.
JJ acknowledges support by NSF grants AST-2009884 and AST-2307983, and NASA grants 80NSSC21K1746, 80NSSC22K0826, 80NSSC24K0940 and NASA XMM-Newton grant 80NSSC22K0799. BL acknowledges support by a National Science Foundation Graduate Research Fellowship under Grant No. DGE-2234667.

AT acknowledges support by NASA 
80NSSC22K0031, 
80NSSC22K0799, 
80NSSC18K0565 
and 80NSSC21K1746 
grants, and by the NSF grants 
AST-2009884, 
AST-2107839, 
AST-1815304, 
AST-1911080, 
AST-2206471, 
AST-2407475, 
AST-2510570, 
OAC-2031997. 
This work was performed in part at the Kavli Institute for Theoretical Physics (KITP) supported by grant NSF PHY-2309135.
This work was performed in part at Aspen Center for Physics, which is supported by National Science Foundation grant PHY-2210452.
This research used resources of the National Energy Research Scientific Computing Center, a DOE Office of Science User Facility supported by the Office of Science of the U.S. Department of Energy under Contract No. DE-AC02-05CH11231 using NERSC allocations m4603 (award NP-ERCAP0029085) and m2401. The computations in this work were, in part, run at facilities supported by the Scientific Computing Core at the Flatiron Institute, a division of the Simons Foundation. An award of computer time was provided by the ASCR Leadership Computing Challenge (ALCC), Innovative and Novel Computational Impact on Theory and Experiment (INCITE), and OLCF Director’s Discretionary Allocation programs under awards PHY129 and AST178.  This research was partially carried out using resources from Calcul Quebec (http://www.calculquebec.ca) and Compute Canada (http://www.computecanada.ca) under RAPI xsp-772-ab (PI: Daryl Haggard). This research also used HPC and visualization resources provided by
the Texas Advanced Computing Center (TACC) at The University
of Texas at Austin, which contributed to our results via the LRAC allocation AST20011 (http://www.tacc.
utexas.edu).

\vspace{5mm}

\appendix
\section{Derivation of latitudinally averaged flux transport equation}
\label{A:deriv_thmean}

Carefully $\theta$-integrating Eq.\eqref{eq:indc_inter} gives
\begin{equation}
\label{eqA:thm_adveq}
\partial_t\tilde{\Phi} + v_\phi\partial_r\tilde{\Phi} = 0,
\end{equation}
where
\begin{equation}
\tilde{\Phi} = \frac{1}{\theta_2-\theta_1}\int_{\theta_1}^{\theta_2}A_\varphi,\mathrm{d}\theta
\end{equation}
is the $\theta$-averaged flux. This definition differs from both the large-scale flux used in this work and the standard MAD literature, where the absolute value is taken:
\begin{equation}
\Phi_{\rm abs}(r,t) = 2\pi \int\limits_0^{\pi} \frac{|\phimean{B^{r}}|}{2} \sqrt{-g},\mathrm{d} \theta.
\label{Eq:pol_flux_abs}
\end{equation}
In practice, we find $\Phi \simeq \Phi_{\rm abs} \simeq \tilde{\Phi}$ in the regions of interest ($r\leq50\,\,r_g$), with multiplicative deviations of at most $\sim 20\%$ (not shown). As shown in Section~\ref{sec:validation}, adopting $\Phi = \tilde{\Phi}$ does not affect accuracy, likely because Eq.\eqref{eq:thm_adveq} is invariant under rescaling $\Phi$ by a constant. For consistency, we use $\Phi$ from Eq.\eqref{Eq:pol_flux} as our definition of the large-scale magnetic flux.

\section{Fits for the turbulent EMF}
\label{A:fits}
In this appendix, we validate our mean-field model, which prescribes the turbulent emf as proportional to the current, with the following form
\begin{equation}
    \mathcal{E}\hatin{\varphi} = \eta J\hatin{\varphi}.
\end{equation}
This prescription allows us to estimate the turbulent resistivity, $\eta$. As shown below, we find excellent agreement between the predictions of the mean-field model and the simulation data.

\begin{figure}
    \centering
    \includegraphics[width=\columnwidth]{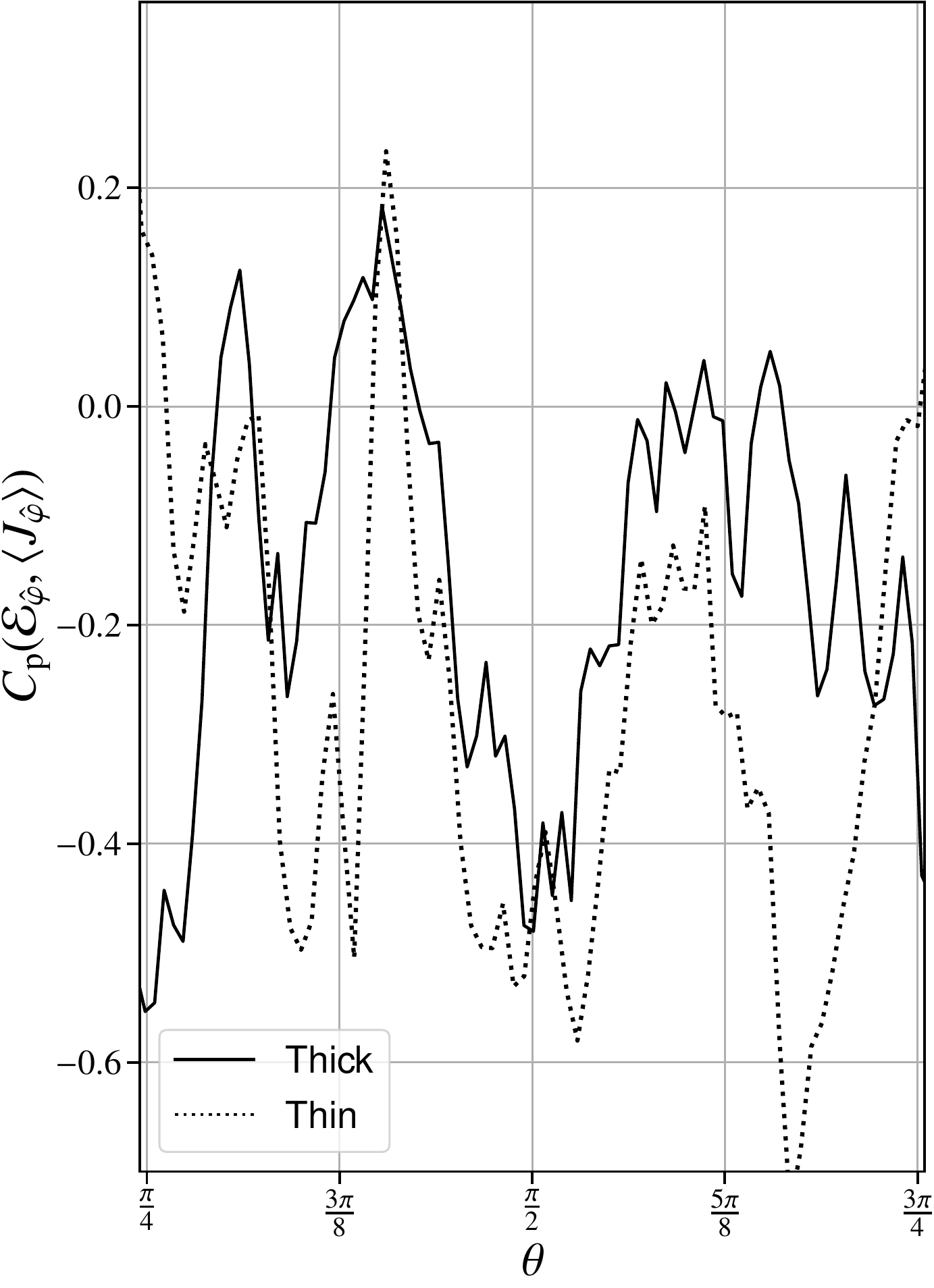}
    \caption{Time-averaged correlation coefficient, $C_{\rm or}(J\hatin{\varphi},\mathcal{E}\hatin{\varphi})$, between the toroidal current and turbulent emf at $r=6\,\,r_g$ for thin and thick disk simulations. The correlation peaks between $-0.4$ and $-0.6$ near the disk midplane in both cases, decreasing above and below it. This indicates that our mean-field estimate of the turbulent resistivity is reliable primarily within $|\theta - \pi/2| \lesssim 0.2$. This correlation is large enough for us to compute $\eta$ with reliable accuracy.}
    \label{fig:corel}
\end{figure}

\begin{figure*}
    \centering
    \includegraphics[width=\textwidth]{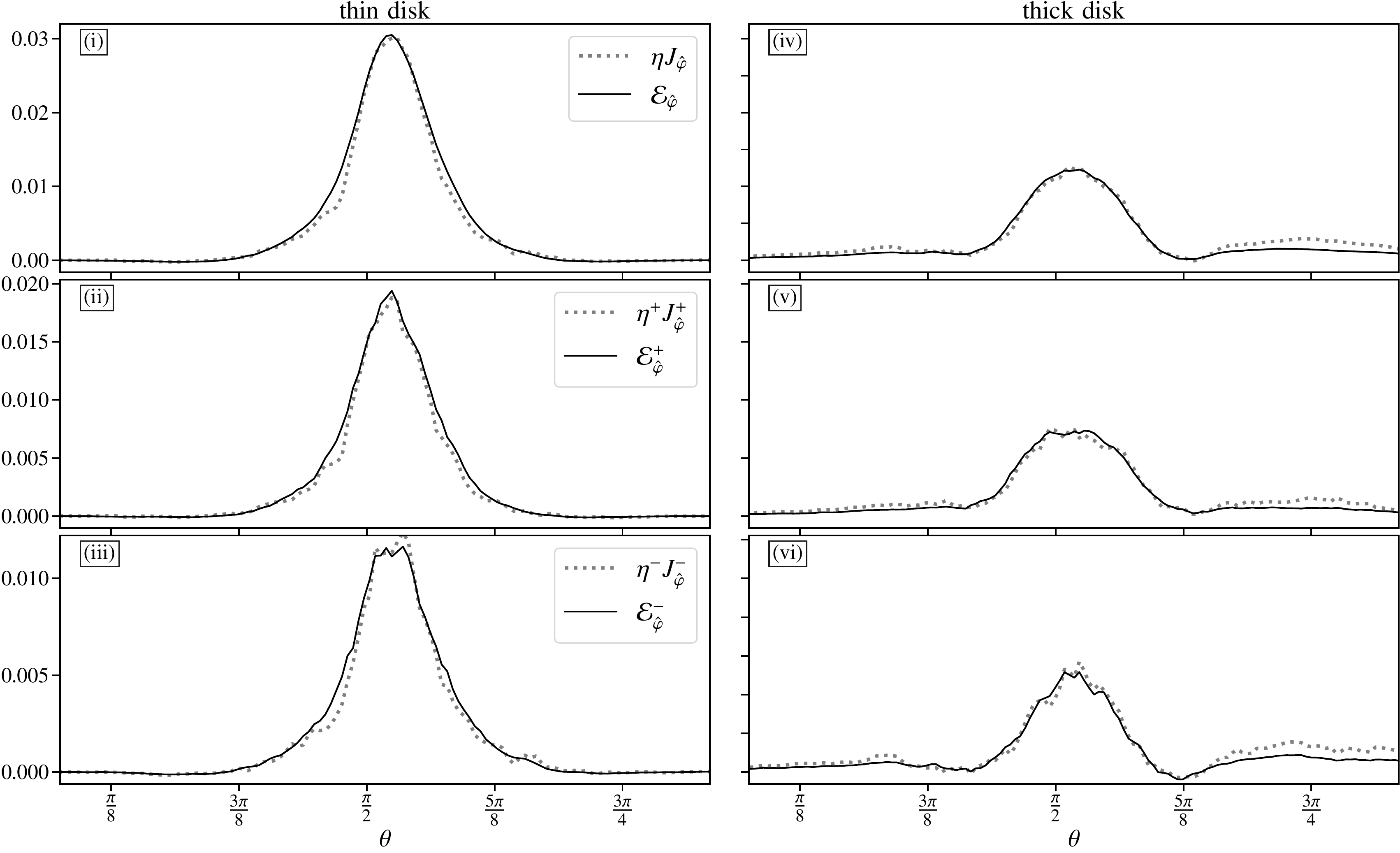}
    \caption{Comparison between the turbulent emf, $\mathcal{E}\hatin{\varphi}$, and its mean-field model, $\eta J\hatin{\varphi}$, at $r=6\,\,r_g$ for thin (i–iii) and thick (iv–vi) disk models. Panels (i) and (iv) show the time-averaged correlation over the entire simulation, demonstrating strong agreement between the model and simulation data. Panels (ii, v) correspond to averaging during diffusive phases ($v_\Phi>0$) noted as $X^{+}$, while panels (iii, vi) correspond to advective phases ($v_\Phi<0$) noted as $X^{-}$. In all cases, the mean-field model reproduces the directly computed emf with excellent accuracy, confirming the robustness of our estimated turbulent resistivity.}
    \label{fig:fits}
\end{figure*}

To build an effective mean-field model for the turbulent diffusivity, we first need to check whether the current, $J\hatin{\varphi}$, and the turbulent emf, $\mathcal{E}\hatin{\varphi}$, are well correlated. To do this, we compute their correlation coefficient defined as
\begin{equation}
\label{eq:cp_app}
C_{\rm or}(X,Y) = \frac{\int\limits^{t_2}_{t_1} X(t')Y(t'),\mathrm{d}t'}{\sqrt{\int\limits^{t_2}_{t_1} X^2(t'),\mathrm{d}t' \int\limits^{t_2}_{t_1} Y^2(t'),\mathrm{d}t'}}.
\end{equation}
We show the time-average of this quantity at $r=6\,\,r_g$ for both thin and thick disk simulations in Fig.~\ref{fig:corel}. We find that it peaks at $-0.4$ to $-0.6$ around the disk midplane in both simulations. The correlation then decreases above and below the midplane for both thin and thick disk models, so we can only trust our estimate of the disk resistivity in regions near the midplane, $|\theta-\pi/2|\lesssim0.2$. We find similar trends at all radii probed in this work, for $r\lesssim30\,\,r_g$.

The correlation is strong enough for us to estimate the resistivity using the following formula:
\begin{equation}
\label{eq:app_eta}
\eta \simeq C,\frac{\int\limits^{t_2}_{t_1} \mathcal{E}\hatin{\varphi}(t')\mean{J\hatin{\varphi}}(t'),\mathrm{d}t'}{\int\limits^{t_2}_{t_1} \mean{J\hatin{\varphi}}^2(t'),\mathrm{d}t'},
\end{equation}
where $C$ is an order-unity coefficient chosen to maximize the goodness of fit. We find that $C=1.4$ for the thin disk model and $C=1.6$ for the thick disk model. In Fig.~\ref{fig:fits}, we show a comparison between the turbulent emf, $\mathcal{E}\hatin{\varphi}$, and its mean-field model, $\eta J\hatin{\varphi}$, evaluated at $r=6\,\,r_g$ for both the thin (i) and thick (iv) disk models. We find remarkable agreement between our mean-field model and the simulation data for both cases. This agreement holds across all radii examined, $r\leq30\,\,r_g$, with the coefficient $C$ always bounded within $1\leq C\leq2.5$.

To understand how the different diffusive coefficients depend on whether the system is currently diffusing or advecting magnetic fields, we define the following averages:
\begin{align}
X^+ &= \int\limits_{t_1}^{t_2}X\mathcal{H}(v_\Phi),\mathrm{d}t,\\
X^- &= \int\limits_{t_1}^{t_2}X\mathcal{H}(-v_\Phi),\mathrm{d}t,
\end{align}
where $\mathcal{H}$ is a Heaviside function. $X^{+}$ represents an average during a diffusive phase, while $X^{-}$ represents an average during an advective phase.
We use these averages to compute $\mathcal{E}\hatin{\varphi}^{\pm}$, $J\hatin{\varphi}^{\pm}$, and $\eta^{\pm}$, making use of Eq.~\ref{eq:app_eta} with the corresponding $\mathcal{H}$. We then compare the turbulent emf averaged during the advective and diffusive phases to the mean-field model averaged in the same way in Fig.~\ref{fig:fits}, evaluated at $r=6\,\,r_g$ for both thin (ii, iii) and thick (v, vi) disk models, and during advective (iii, vi) and diffusive (ii, v) phases. As with the full averages, we find excellent agreement between the mean-field model and the directly calculated emf in both advective and diffusive phases for both thin and thick disks. Furthermore, the coefficient $C$ remains unchanged between advective and diffusive phases, lending additional confidence to our methodology.

We conclude that our estimates of $\eta$ presented in Section~\ref{sec:eff_res} are accurate and fully consistent with the simulation data. They can therefore be confidently used in our interpretations within this work and may be extended for use in simplified models \citep{ferreira_magnetized_1995,jacquemin-ide_magnetically-driven_2019,zimniak_influence_2024}.

\bibliographystyle{aasjournal}
\bibliography{biblio}

\begin{thebibliography}{}
\expandafter\ifx\csname natexlab\endcsname\relax\def\natexlab#1{#1}\fi
\providecommand{\url}[1]{\href{#1}{#1}}
\providecommand{\dodoi}[1]{doi:~\href{http://doi.org/#1}{\nolinkurl{#1}}}
\providecommand{\doeprint}[1]{\href{http://ascl.net/#1}{\nolinkurl{http://ascl.net/#1}}}
\providecommand{\doarXiv}[1]{\href{https://arxiv.org/abs/#1}{\nolinkurl{https://arxiv.org/abs/#1}}}

\bibitem[{Abuter {et~al.}(2018)Abuter, Amorim, Bauböck, Berger, Bonnet, Brandner, Clénet, Foresto, Zeeuw, Deen, Dexter, Duvert, Eckart, Eisenhauer, Schreiber, Garcia, Gao, Gendron, Genzel, Gillessen, Guajardo, Habibi, Haubois, Henning, Hippler, Horrobin, Huber, Jiménez-Rosales, Jocou, Kervella, Lacour, Lapeyrère, Lazareff, Bouquin, Léna, Lippa, Ott, Panduro, Paumard, Perraut, Perrin, Pfuhl, Plewa, Rabien, Rodríguez-Coira, Rousset, Sternberg, Straub, Straubmeier, Sturm, Tacconi, Vincent, Fellenberg, Waisberg, Widmann, Wieprecht, Wiezorrek, Woillez, \& Yazici}]{abuter_detection_2018}
Abuter, R., Amorim, A., Bauböck, M., {et~al.} 2018, Astronomy \& Astrophysics, 618, L10, \dodoi{10.1051/0004-6361/201834294}

\bibitem[{Aharonian {et~al.}(2006)Aharonian, Akhperjanian, Bazer-Bachi, Beilicke, Benbow, Berge, Bernlöhr, Boisson, Bolz, Borrel, Braun, Brown, Bühler, Büsching, Carrigan, Chadwick, Chounet, Coignet, Cornils, Costamante, Degrange, Dickinson, Djannati-Ataï, Drury, Dubus, Egberts, Emmanoulopoulos, Espigat, Feinstein, Ferrero, Fiasson, Fontaine, Funk, Funk, Füßling, Gallant, Giebels, Glicenstein, Goret, Hadjichristidis, Hauser, Hauser, Heinzelmann, Henri, Hermann, Hinton, Hoffmann, Hofmann, Holleran, Hoppe, Horns, Jacholkowska, de~Jager, Kendziorra, Kerschhaggl, Khélifi, Komin, Konopelko, Kosack, Lamanna, Latham, Le~Gallou, Lemière, Lemoine-Goumard, Lenain, Lohse, Martin, Martineau-Huynh, Marcowith, Masterson, Maurin, McComb, Moulin, de~Naurois, Nedbal, Nolan, Noutsos, Orford, Osborne, Ouchrif, Panter, Pelletier, Pita, Pühlhofer, Punch, Ranchon, Raubenheimer, Raue, Rayner, Reimer, Ripken, Rob, Rolland, Rosier-Lees, Rowell, Sahakian, Santangelo, Saugé, Schlenker, Schlickeiser, Schröder, Schwanke,
  Schwarzburg, Schwemmer, Shalchi, Sol, Spangler, Spanier, Steenkamp, Stegmann, Superina, Tam, Tavernet, Terrier, Tluczykont, van Eldik, Vasileiadis, Venter, Vialle, Vincent, Völk, Wagner, \& Ward}]{aharonian_fast_2006}
Aharonian, F., Akhperjanian, A.~G., Bazer-Bachi, A.~R., {et~al.} 2006, Science, 314, 1424, \dodoi{10.1126/science.1134408}

\bibitem[{Albert {et~al.}(2007)Albert, Aliu, Anderhub, Antoranz, Armada, Baixeras, Barrio, Bartko, Bastieri, Becker, Bednarek, Berger, Bigongiari, Biland, Bock, Bordas, Bosch-Ramon, Bretz, Britvitch, Camara, Carmona, Chilingarian, Coarasa, Commichau, Contreras, Cortina, Costado, Curtef, Danielyan, Dazzi, Angelis, Delgado, Reyes, Lotto, Domingo-Santamaría, Dorner, Doro, Errando, Fagiolini, Ferenc, Fernández, Firpo, Flix, Fonseca, Font, Fuchs, Galante, García-López, Garczarczyk, Gaug, Giller, Goebel, Hakobyan, Hayashida, Hengstebeck, Herrero, Höhne, Hose, Hrupec, Hsu, Jacon, Jogler, Kosyra, Kranich, Kritzer, Laille, Lindfors, Lombardi, Longo, López, López, Lorenz, Majumdar, Maneva, Mannheim, Mansutti, Mariotti, Martínez, Mazin, Merck, Meucci, Meyer, Miranda, Mirzoyan, Mizobuchi, Moralejo, Nieto, Nilsson, Ninkovic, Oña-Wilhelmi, Otte, Oya, Paneque, Panniello, Paoletti, Paredes, Pasanen, Pascoli, Pauss, Pegna, Persic, Peruzzo, Piccioli, Prandini, Puchades, Raymers, Rhode, Ribó, Rico, Rissi, Robert,
  Rügamer, Saggion, Saito, Sánchez, Sartori, Scalzotto, Scapin, Schmitt, Schweizer, Shayduk, Shinozaki, Shore, Sidro, Sillanpää, Sobczynska, Stamerra, Stark, Takalo, Tavecchio, Temnikov, Tescaro, Teshima, Torres, Turini, Vankov, Vitale, Wagner, Wibig, Wittek, Zandanel, Zanin, \& Zapatero}]{albert_variable_2007}
Albert, J., Aliu, E., Anderhub, H., {et~al.} 2007, The Astrophysical Journal, 669, 862, \dodoi{10.1086/521382}

\bibitem[{Aleksić {et~al.}(2014)Aleksić, Ansoldi, Antonelli, Antoranz, Babic, Bangale, Barrio, González, Bednarek, Bernardini, Biasuzzi, Biland, Blanch, Bonnefoy, Bonnoli, Borracci, Bretz, Carmona, Carosi, Colin, Colombo, Contreras, Cortina, Covino, Da~Vela, Dazzi, De~Angelis, De~Caneva, De~Lotto, Wilhelmi, Mendez, Prester, Dorner, Doro, Einecke, Eisenacher, Elsaesser, Fonseca, Font, Frantzen, Fruck, Galindo, López, Garczarczyk, Terrats, Gaug, Godinović, Muñoz, Gozzini, Hadasch, Hanabata, Hayashida, Herrera, Hildebrand, Hose, Hrupec, Idec, Kadenius, Kellermann, Kodani, Konno, Krause, Kubo, Kushida, La~Barbera, Lelas, Lewandowska, Lindfors, Lombardi, Longo, López, López-Coto, López-Oramas, Lorenz, Lozano, Makariev, Mallot, Maneva, Mankuzhiyil, Mannheim, Maraschi, Marcote, Mariotti, Martínez, Mazin, Menzel, Miranda, Mirzoyan, Moralejo, Munar-Adrover, Nakajima, Niedzwiecki, Nilsson, Nishijima, Noda, Orito, Overkemping, Paiano, Palatiello, Paneque, Paoletti, Paredes, Paredes-Fortuny, Persic, Poutanen,
  Moroni, Prandini, Puljak, Reinthal, Rhode, Ribó, Rico, Garcia, Rügamer, Saito, Saito, Satalecka, Scalzotto, Scapin, Schultz, Schweizer, Shore, Sillanpää, Sitarek, Snidaric, Sobczynska, Spanier, Stamatescu, Stamerra, Steinbring, Storz, Strzys, Takalo, Takami, Tavecchio, Temnikov, Terzić, Tescaro, Teshima, Thaele, Tibolla, Torres, Toyama, Treves, Uellenbeck, Vogler, Zanin, Kadler, Schulz, Ros, Bach, Krauß, \& Wilms}]{aleksic_black_2014}
Aleksić, J., Ansoldi, S., Antonelli, L.~A., {et~al.} 2014, Science, 346, 1080, \dodoi{10.1126/science.1256183}

\bibitem[{Aliu {et~al.}(2012)Aliu, Arlen, Aune, Beilicke, Benbow, Bouvier, Bradbury, Buckley, Bugaev, Byrum, Cannon, Cesarini, Ciupik, Collins-Hughes, Connolly, Cui, Dickherber, Duke, Errando, Falcone, Finl @article{ripperda_black_2022, title = {Black {Hole} {Flares}: {Ejection} of {Accreted} {Magnetic} {Flux} through {3D} {Plasmoid}-mediated {Reconnection}}, volume = {924}, issn = {0004-637X}, shorttitle = {Black {Hole} {Flares}}, url = {https://ui.adsabs.harvard.edu/abs/2022ApJ...924L..32R}, doi = {10.3847/2041-8213/ac46a1}, abstract = {Magnetic reconnection can power bright, rapid flares originating from the inner magnetosphere of accreting black holes. We conduct extremely high-resolution (5376 × 2304 × 2304 cells) general-relativistic magnetohydrodynamics simulations, capturing plasmoid-mediated reconnection in a 3D magnetically arrested disk for the first time. We show that an equatorial, plasmoid-unstable current sheet forms in a transient, nonaxisymmetric, low-density magnetosphere within the inner
  few Schwarzschild radii. Magnetic flux bundles escape from the event horizon through reconnection at the universal plasmoid-mediated rate in this current sheet. The reconnection feeds on the highly magnetized plasma in the jets and heats the plasma that ends up trapped in flux bundles to temperatures proportional to the jet's magnetization. The escaped flux bundles can complete a full orbit as low-density hot spots, consistent with Sgr A* observations by the GRAVITY interferometer. Reconnection near the horizon produces sufficiently energetic plasma to explain flares from accreting black holes, such as the TeV emission observed from M87. The drop in the mass accretion rate during the flare and the resulting low-density magnetosphere make it easier for very-high-energy photons produced by reconnection-accelerated particles to escape. The extreme-resolution results in a converged plasmoid-mediated reconnection rate that directly determines the timescales and properties of the flare.}, urldate = {2024-03-14},
  journal = {The Astrophysical Journal}, author = {Ripperda, B. and Liska, M. and Chatterjee, K. and Musoke, G. and Philippov, A. A. and Markoff, S. B. and Tchekhovskoy, A. and Younsi, Z.}, month = jan, year = {2022}, note = {ADS Bibcode: 2022ApJ...924L..32R}, keywords = {Astrophysics - High Energy Astrophysical Phenomena, Physics - Plasma Physics, General Relativity and Quantum Cosmology, 162, 1964, 641, 1261, 739}, pages = {L32}, file = {Full Text PDF:/home/jacquejo/Zotero/storage/RHXQLKUA/Ripperda et al. - 2022 - Black Hole Flares Ejection of Accreted Magnetic F.pdf:application/pdf}, }~@article{zhdankin_particle_2023, title = {Particle acceleration by magnetic {Rayleigh}-{Taylor} instability: {Mechanism} for flares in black hole accretion flows}, volume = {5}, issn = {2643-1564}, shorttitle = {Particle acceleration by magnetic {Rayleigh}-{Taylor} instability}, url = {https://link.aps.org/doi/10.1103/PhysRevResearch.5.043023}, doi = {10.1103/PhysRevResearch.5.043023}, language = {en}, number = {4}, urldate
  = {2025-05-28}, journal = {Physical Review Research}, author = {Zhdankin, Vladimir and Ripperda, Bart and Philippov, Alexander A.}, month = oct, year = {2023}, pages = {043023}, file = {Zhdankin et al. - 2023 - Particle acceleration by magnetic Rayleigh-Taylor .pdf:/home/jacquejo/Zotero/storage/72SKWNIY/Zhdankin et al. - 2023 - Particle acceleration by magnetic Rayleigh-Taylor .pdf:application/pdf}, } ey, Finnegan, Fortson, Furniss, Galante, Gall, Godambe, Griffin, Grube, Guenette, Gyuk, Hanna, Holder, Huan, Hughes, Hui, Humensky, Imran, Kaaret, Karlsson, Kertzman, Kieda, Krawczynski, Krennrich, Lang, LeBohec, Madhavan, Maier, Majumdar, McArthur, McCann, Moriarty, Mukherjee, Nuñez, Ong, Orr, Otte, Park, Perkins, Pichel, Pohl, Prokoph, Quinn, Ragan, Reyes, Reynolds, Roache, Rose, Ruppel, Saxon, Schroedter, Sembroski, Şentürk, Skole, Staszak, Tešić, Theiling, Thibadeau, Tsurusaki, Tyler, Varlotta, Vassiliev, Vincent, Vivier, Wakely, Ward, Weekes, Weinstein, Weisgarber, Williams, \&
  Zitzer}]{aliu_veritas_2012}
Aliu, E., Arlen, T., Aune, T., {et~al.} 2012, The Astrophysical Journal, 746, 141, \dodoi{10.1088/0004-637X/746/2/141}

\bibitem[{Avara {et~al.}(2016)Avara, McKinney, \& Reynolds}]{avara_efficiency_2016}
Avara, M.~J., McKinney, J.~C., \& Reynolds, C.~S. 2016, Monthly Notices of the Royal Astronomical Society, 462, 636, \dodoi{10.1093/mnras/stw1643}

\bibitem[{Baganoff {et~al.}(2001)Baganoff, Bautz, Brandt, Chartas, Feigelson, Garmire, Maeda, Morris, Ricker, Townsley, \& Walter}]{baganoff_rapid_2001}
Baganoff, F.~K., Bautz, M.~W., Brandt, W.~N., {et~al.} 2001, Nature, 413, 45, \dodoi{10.1038/35092510}

\bibitem[{Balbus \& Hawley(1991)}]{balbus_powerful_1991}
Balbus, S.~A., \& Hawley, J.~F. 1991, The Astrophysical Journal, 376, 214, \dodoi{10.1086/170270}

\bibitem[{Balbus \& Hawley(1998)}]{balbus_instability_1998}
---. 1998, Reviews of Modern Physics, 70, 1, \dodoi{10.1103/RevModPhys.70.1}

\bibitem[{Beckwith {et~al.}(2009)Beckwith, Hawley, \& Krolik}]{beckwith_transport_2009}
Beckwith, K., Hawley, J.~F., \& Krolik, J.~H. 2009, The Astrophysical Journal, 707, 428, \dodoi{10.1088/0004-637X/707/1/428}

\bibitem[{Begelman(2024)}]{begelman_simple_2024}
Begelman, M.~C. 2024, Monthly Notices of the Royal Astronomical Society, 534, 3144, \dodoi{10.1093/mnras/stae2305}

\bibitem[{Begelman \& Armitage(2023)}]{begelman_saturation_2023}
Begelman, M.~C., \& Armitage, P.~J. 2023, Monthly Notices of the Royal Astronomical Society, 521, 5952, \dodoi{10.1093/mnras/stad914}

\bibitem[{Begelman {et~al.}(2022)Begelman, Scepi, \& Dexter}]{begelman_what_2022}
Begelman, M.~C., Scepi, N., \& Dexter, J. 2022, Monthly Notices of the Royal Astronomical Society, 511, 2040, \dodoi{10.1093/mnras/stab3790}

\bibitem[{Blandford \& Payne(1982)}]{blandford_hydromagnetic_1982}
Blandford, R.~D., \& Payne, D.~G. 1982, Monthly Notices of the Royal Astronomical Society, 199, 883, \dodoi{10.1093/mnras/199.4.883}

\bibitem[{Blandford \& Znajek(1977)}]{blandford_electromagnetic_1977}
Blandford, R.~D., \& Znajek, R.~L. 1977, Monthly Notices of the Royal Astronomical Society, 179, 433, \dodoi{10.1093/mnras/179.3.433}

\bibitem[{Brughmans \& Keppens(2025)}]{brughmans_visual_2025}
Brughmans, N., \& Keppens, R. 2025, Monthly Notices of the Royal Astronomical Society, 542, 1347, \dodoi{10.1093/mnras/staf1265}

\bibitem[{Brughmans {et~al.}(2024)Brughmans, Keppens, \& Goedbloed}]{brughmans_parametric_2024}
Brughmans, N., Keppens, R., \& Goedbloed, H. 2024, The Astrophysical Journal, 968, 19, \dodoi{10.3847/1538-4357/ad3d52}

\bibitem[{Burrows {et~al.}(2011)Burrows, Kennea, Ghisellini, Mangano, Zhang, Page, Eracleous, Romano, Sakamoto, Falcone, Osborne, Campana, Beardmore, Breeveld, Chester, Corbet, Covino, Cummings, D'Avanzo, D'Elia, Esposito, Evans, Fugazza, Gelbord, Hiroi, Holland, Huang, Im, Israel, Jeon, Jeon, Jun, Kawai, Kim, Krimm, Marshall, {P. Mészáros}, Negoro, Omodei, Park, Perkins, Sugizaki, Sung, Tagliaferri, Troja, Ueda, Urata, Usui, Antonelli, Barthelmy, Cusumano, Giommi, Melandri, Perri, Racusin, Sbarufatti, Siegel, \& Gehrels}]{burrows_relativistic_2011}
Burrows, D.~N., Kennea, J.~A., Ghisellini, G., {et~al.} 2011, Nature, 476, 421, \dodoi{10.1038/nature10374}

\bibitem[{{Chakrabarti}(1985)}]{chakrabarti_1985}
{Chakrabarti}, S.~K. 1985, \apj, 288, 1, \dodoi{10.1086/162755}

\bibitem[{Chatterjee \& Narayan(2022)}]{chatterjee_flux_2022}
Chatterjee, K., \& Narayan, R. 2022, The Astrophysical Journal, 941, 30, \dodoi{10.3847/1538-4357/ac9d97}

\bibitem[{Corbel {et~al.}(2013)Corbel, Coriat, Brocksopp, Tzioumis, Fender, Tomsick, Buxton, \& Bailyn}]{corbel_universal_2013}
Corbel, S., Coriat, M., Brocksopp, C., {et~al.} 2013, Monthly Notices of the Royal Astronomical Society, 428, 2500, \dodoi{10.1093/mnras/sts215}

\bibitem[{Corbel {et~al.}(2003)Corbel, Nowak, Fender, Tzioumis, \& Markoff}]{corbel_radio/x-ray_2003}
Corbel, S., Nowak, M.~A., Fender, R.~P., Tzioumis, A.~K., \& Markoff, S. 2003, Astronomy and Astrophysics, 400, 1007, \dodoi{10.1051/0004-6361:20030090}

\bibitem[{Das {et~al.}(2018)Das, Begelman, \& Lesur}]{das_instability_2018}
Das, U., Begelman, M.~C., \& Lesur, G. 2018, Monthly Notices of the Royal Astronomical Society, 473, 2791, \dodoi{10.1093/mnras/stx2518}

\bibitem[{Dexter {et~al.}(2020)Dexter, Tchekhovskoy, Jiménez-Rosales, Ressler, Bauböck, Dallilar, de Zeeuw, Eisenhauer, von Fellenberg, Gao, Genzel, Gillessen, Habibi, Ott, Stadler, Straub, \& Widmann}]{dexter_sgr_2020}
Dexter, J., Tchekhovskoy, A., Jiménez-Rosales, A., {et~al.} 2020, Monthly Notices of the Royal Astronomical Society, 497, 4999, \dodoi{10.1093/mnras/staa2288}

\bibitem[{Done {et~al.}(2007)Done, Gierliński, \& Kubota}]{done_modelling_2007}
Done, C., Gierliński, M., \& Kubota, A. 2007, Astronomy and Astrophysics Review, 15, 1, \dodoi{10.1007/s00159-007-0006-1}

\bibitem[{EHT~Collaboration {et~al.}(2021)EHT~Collaboration, Akiyama, Algaba, Alberdi, Alef, Anantua, Asada, Azulay, Baczko, Ball, Baloković, Barrett, Benson, Bintley, Blackburn, Blundell, Boland, Bouman, Bower, Boyce, Bremer, Brinkerink, Brissenden, Britzen, Broderick, Broguiere, Bronzwaer, Byun, Carlstrom, Chael, Chan, Chatterjee, Chatterjee, Chen, Chen, Chesler, Cho, Christian, Conway, Cordes, Crawford, Crew, Cruz-Osorio, Cui, Davelaar, Laurentis, Deane, Dempsey, Desvignes, Dexter, Doeleman, Eatough, Falcke, Farah, Fish, Fomalont, Ford, Fraga-Encinas, Friberg, Fromm, Fuentes, Galison, Gammie, García, Gelles, Gentaz, Georgiev, Goddi, Gold, Gómez, Gómez-Ruiz, Gu, Gurwell, Hada, Haggard, Hecht, Hesper, Himwich, Ho, Ho, Honma, Huang, Huang, Hughes, Ikeda, Inoue, Issaoun, James, Jannuzi, Janssen, Jeter, Jiang, Jimenez-Rosales, Johnson, Jorstad, Jung, Karami, Karuppusamy, Kawashima, Keating, Kettenis, Kim, Kim, Kim, Kim, Kino, Koay, Kofuji, Koch, Koyama, Kramer, Kramer, Krichbaum, Kuo, Lauer, Lee, Levis, Li,
  Li, Lindqvist, Lico, Lindahl, Liu, Liu, Liuzzo, Lo, Lobanov, Loinard, Lonsdale, Lu, MacDonald, Mao, Marchili, Markoff, Marrone, Marscher, Martí-Vidal, Matsushita, Matthews, Medeiros, Menten, Mizuno, Mizuno, Moran, Moriyama, Moscibrodzka, Müller, Musoke, Mejías, Michalik, Nadolski, Nagai, Nagar, Nakamura, Narayan, Narayanan, Natarajan, Nathanail, Neilsen, Neri, Ni, Noutsos, Nowak, Okino, Olivares, Ortiz-León, Oyama, Özel, Palumbo, Park, Patel, Pen, Pesce, Piétu, Plambeck, PopStefanija, Porth, Pötzl, Prather, Preciado-López, Psaltis, Pu, Ramakrishnan, Rao, Rawlings, Raymond, Rezzolla, Ricarte, Ripperda, Roelofs, Rogers, Ros, Rose, Roshanineshat, Rottmann, Roy, Ruszczyk, Rygl, Sánchez, Sánchez-Arguelles, Sasada, Savolainen, Schloerb, Schuster, Shao, Shen, Small, Sohn, SooHoo, Sun, Tazaki, Tetarenko, Tiede, Tilanus, Titus, Toma, Torne, Trent, Traianou, Trippe, Bemmel, Langevelde, Rossum, Wagner, Ward-Thompson, Wardle, Weintroub, Wex, Wharton, Wielgus, Wong, Wu, Yoon, Young, Young, Younsi, Yuan, Yuan,
  Zensus, Zhao, Zhao, \& Collaboration}]{collaboration_first_2021}
EHT~Collaboration, T. E. H.~T., Akiyama, K., Algaba, J.~C., {et~al.} 2021, The Astrophysical Journal Letters, 910, L13, \dodoi{10.3847/2041-8213/abe4de}

\bibitem[{Ferreira \& Pelletier(1995)}]{ferreira_magnetized_1995}
Ferreira, J., \& Pelletier, G. 1995, Astronomy and Astrophysics, 295, 807.
\newblock \url{http://cdsads.u-strasbg.fr/abs/1995A%26A...295..807F}

\bibitem[{Ferreira {et~al.}(2006)Ferreira, Petrucci, Henri, Saugé, \& Pelletier}]{ferreira_unified_2006}
Ferreira, J., Petrucci, P.-O., Henri, G., Saugé, L., \& Pelletier, G. 2006, Astronomy and Astrophysics, 447, 813, \dodoi{10.1051/0004-6361:20052689}

\bibitem[{Fishbone \& Moncrief(1976)}]{fishbone_relativistic_1976}
Fishbone, L.~G., \& Moncrief, V. 1976, The Astrophysical Journal, 207, 962, \dodoi{10.1086/154565}

\bibitem[{Fromang \& Stone(2009)}]{fromang_turbulent_2009}
Fromang, S., \& Stone, J.~M. 2009, Astronomy and Astrophysics, 507, 19, \dodoi{10.1051/0004-6361/200912752}

\bibitem[{Gammie {et~al.}(2003)Gammie, McKinney, \& Tóth}]{gammie_harm_2003}
Gammie, C.~F., McKinney, J.~C., \& Tóth, G. 2003, The Astrophysical Journal, 589, 444, \dodoi{10.1086/374594}

\bibitem[{Genzel {et~al.}(2003)Genzel, Schödel, Ott, Eckart, Alexander, Lacombe, Rouan, \& Aschenbach}]{genzel_near-infrared_2003}
Genzel, R., Schödel, R., Ott, T., {et~al.} 2003, Nature, 425, 934, \dodoi{10.1038/nature02065}

\bibitem[{Goedbloed \& Keppens(2022)}]{goedbloed_super-alfvenic_2022}
Goedbloed, H., \& Keppens, R. 2022, The Astrophysical Journal Supplement Series, 259, 65, \dodoi{10.3847/1538-4365/ac573c}

\bibitem[{Guan \& Gammie(2009)}]{guan_turbulent_2009}
Guan, X., \& Gammie, C.~F. 2009, The Astrophysical Journal, 697, 1901, \dodoi{10.1088/0004-637X/697/2/1901}

\bibitem[{Guilet \& Ogilvie(2012)}]{guilet_transport_2012}
Guilet, J., \& Ogilvie, G.~I. 2012, Monthly Notices of the Royal Astronomical Society, 424, 2097, \dodoi{10.1111/j.1365-2966.2012.21361.x}

\bibitem[{Guilet \& Ogilvie(2013)}]{guilet_transport_2013}
---. 2013, Monthly Notices of the Royal Astronomical Society, 430, 822, \dodoi{10.1093/mnras/sts551}

\bibitem[{Hakobyan {et~al.}(2023)Hakobyan, Ripperda, \& Philippov}]{hakobyan_radiative_2023}
Hakobyan, H., Ripperda, B., \& Philippov, A.~A. 2023, The Astrophysical Journal Letters, 943, L29, \dodoi{10.3847/2041-8213/acb264}

\bibitem[{Hawley {et~al.}(1995)Hawley, Gammie, \& Balbus}]{hawley_local_1995}
Hawley, J.~F., Gammie, C.~F., \& Balbus, S.~A. 1995, The Astrophysical Journal, 440, 742, \dodoi{10.1086/175311}

\bibitem[{Jacquemin-Ide {et~al.}(2019)Jacquemin-Ide, Ferreira, \& Lesur}]{jacquemin-ide_magnetically-driven_2019}
Jacquemin-Ide, J., Ferreira, J., \& Lesur, G. 2019, Monthly Notices of the Royal Astronomical Society, 490, 3112, \dodoi{10.1093/mnras/stz2749}

\bibitem[{Jacquemin-Ide {et~al.}(2021)Jacquemin-Ide, Lesur, \& Ferreira}]{jacquemin-ide_magnetic_2021}
Jacquemin-Ide, J., Lesur, G., \& Ferreira, J. 2021, Astronomy \& Astrophysics, 647, A192, \dodoi{10.1051/0004-6361/202039322}

\bibitem[{Jacquemin-Ide {et~al.}(2024)Jacquemin-Ide, Rincon, Tchekhovskoy, \& Liska}]{jacquemin-ide_magnetorotational_2024}
Jacquemin-Ide, J., Rincon, F., Tchekhovskoy, A., \& Liska, M. 2024, Monthly Notices of the Royal Astronomical Society, 532, 1522, \dodoi{10.1093/mnras/stae1538}

\bibitem[{Jiménez-Rosales {et~al.}(2020)Jiménez-Rosales, Dexter, Widmann, Bauböck, Abuter, Amorim, Berger, Bonnet, Brandner, Clénet, Zeeuw, Eckart, Eisenhauer, Schreiber, Garcia, Gao, Gendron, Genzel, Gillessen, Habibi, Haubois, Heißel, Henning, Hippler, Horrobin, Jochum, Jocou, Kaufer, Kervella, Lacour, Lapeyrère, Bouquin, Léna, Nowak, Ott, Paumard, Perraut, Perrin, Pfuhl, Rodríguez-Coira, Shangguan, Scheithauer, Stadler, Straub, Straubmeier, Sturm, Tacconi, Vincent, Fellenberg, Waisberg, Wieprecht, Wiezorrek, Woillez, Yazici, \& Zins}]{jimenez-rosales_dynamically_2020}
Jiménez-Rosales, A., Dexter, J., Widmann, F., {et~al.} 2020, Astronomy \& Astrophysics, 643, A56, \dodoi{10.1051/0004-6361/202038283}

\bibitem[{Kaaz {et~al.}(2025)Kaaz, Liska, Tchekhovskoy, Hopkins, \& Jacquemin-Ide}]{kaaz_h-amr_2025}
Kaaz, N., Liska, M., Tchekhovskoy, A., Hopkins, P.~F., \& Jacquemin-Ide, J. 2025, The Astrophysical Journal, 979, 248, \dodoi{10.3847/1538-4357/ad9a86}

\bibitem[{Latter {et~al.}(2010)Latter, Fromang, \& Gressel}]{latter_mri_2010}
Latter, H.~N., Fromang, S., \& Gressel, O. 2010, Monthly Notices of the Royal Astronomical Society, 406, 848, \dodoi{10.1111/j.1365-2966.2010.16759.x}

\bibitem[{Lesur \& Longaretti(2009)}]{lesur_turbulent_2009}
Lesur, G., \& Longaretti, P.-Y. 2009, Astronomy \& Astrophysics, 504, 309, \dodoi{10.1051/0004-6361/200912272}

\bibitem[{Li \& Cao(2021)}]{li_large-scale_2021}
Li, J.-w., \& Cao, X. 2021, The Astrophysical Journal, 909, 158, \dodoi{10.3847/1538-4357/abe125}

\bibitem[{Liska {et~al.}(2020)Liska, Tchekhovskoy, \& Quataert}]{liska_large-scale_2020}
Liska, M., Tchekhovskoy, A., \& Quataert, E. 2020, Monthly Notices of the Royal Astronomical Society, 494, 3656, \dodoi{10.1093/mnras/staa955}

\bibitem[{Liska {et~al.}(2024)Liska, Kaaz, Chatterjee, Emami, \& Musoke}]{liska_magnetic_2024}
Liska, M. T.~P., Kaaz, N., Chatterjee, K., Emami, R., \& Musoke, G. 2024, The Astrophysical Journal, 966, 47, \dodoi{10.3847/1538-4357/ad344a}

\bibitem[{Liska {et~al.}(2022)Liska, Musoke, Tchekhovskoy, Porth, \& Beloborodov}]{liska_formation_2022}
Liska, M. T.~P., Musoke, G., Tchekhovskoy, A., Porth, O., \& Beloborodov, A.~M. 2022, The Astrophysical Journal, 935, L1, \dodoi{10.3847/2041-8213/ac84db}

\bibitem[{{Liska} {et~al.}(2022){Liska}, {Chatterjee}, {Issa}, {Yoon}, {Kaaz}, {Tchekhovskoy}, {van Eijnatten}, {Musoke}, {Hesp}, {Rohoza}, {Markoff}, {Ingram}, \& {van der Klis}}]{Liska2022}
{Liska}, M.~T.~P., {Chatterjee}, K., {Issa}, D., {et~al.} 2022, \apjs, 263, 26, \dodoi{10.3847/1538-4365/ac9966}

\bibitem[{Liska {et~al.}(2022)Liska, Chatterjee, Issa, Yoon, Kaaz, Tchekhovskoy, van Eijnatten, Musoke, Hesp, Rohoza, Markoff, Ingram, \& van~der Klis}]{liska_h-amr_2022}
Liska, M. T.~P., Chatterjee, K., Issa, D., {et~al.} 2022, The Astrophysical Journal Supplement Series, 263, 26, \dodoi{10.3847/1538-4365/ac9966}

\bibitem[{Lowell {et~al.}(2025)Lowell, Jacquemin-Ide, Liska, \& Tchekhovskoy}]{lowell_evidence_2025}
Lowell, B., Jacquemin-Ide, J., Liska, M., \& Tchekhovskoy, A. 2025, Evidence for {Low} {Universal} {Equilibrium} {Black} {Hole} {Spin} in {Luminous} {Magnetically} {Arrested} {Disks}, \dodoi{10.48550/arXiv.2502.17559}

\bibitem[{Lubow {et~al.}(1994)Lubow, Papaloizou, \& Pringle}]{lubow_magnetic_1994}
Lubow, S.~H., Papaloizou, J. C.~B., \& Pringle, J.~E. 1994, Monthly Notices of the Royal Astronomical Society, 267, 235, \dodoi{10.1093/mnras/267.2.235}

\bibitem[{Lubow \& Spruit(1995)}]{lubow_magnetic_1995}
Lubow, S.~H., \& Spruit, H.~C. 1995, The Astrophysical Journal, 445, 337, \dodoi{10.1086/175698}

\bibitem[{Manikantan {et~al.}(2023)Manikantan, Kaaz, Jacquemin-Ide, Musoke, Chatterjee, Liska, \& Tchekhovskoy}]{manikantan_winds_2023}
Manikantan, V., Kaaz, N., Jacquemin-Ide, J., {et~al.} 2023,  arXiv, \dodoi{10.48550/arXiv.2310.11490}

\bibitem[{Marcel {et~al.}(2019)Marcel, Ferreira, Clavel, Petrucci, Malzac, Corbel, Rodriguez, Belmont, Coriat, Henri, \& Cangemi}]{marcel_unified_2019}
Marcel, G., Ferreira, J., Clavel, M., {et~al.} 2019, Astronomy \& Astrophysics, 626, A115, \dodoi{10.1051/0004-6361/201935060}

\bibitem[{Marrone {et~al.}(2008)Marrone, Baganoff, Morris, Moran, Ghez, Hornstein, Dowell, Muñoz, Bautz, Ricker, Brandt, Garmire, Lu, Matthews, Zhao, Rao, \& Bower}]{marrone_x-ray_2008}
Marrone, D.~P., Baganoff, F.~K., Morris, M.~R., {et~al.} 2008, The Astrophysical Journal, 682, 373, \dodoi{10.1086/588806}

\bibitem[{McKinney {et~al.}(2012)McKinney, Tchekhovskoy, \& Blandford}]{mckinney_general_2012}
McKinney, J.~C., Tchekhovskoy, A., \& Blandford, R.~D. 2012, Monthly Notices of the Royal Astronomical Society, 423, 3083, \dodoi{10.1111/j.1365-2966.2012.21074.x}

\bibitem[{Mishra {et~al.}(2020)Mishra, Begelman, Armitage, \& Simon}]{mishra_strongly_2020}
Mishra, B., Begelman, M.~C., Armitage, P.~J., \& Simon, J.~B. 2020, Monthly Notices of the Royal Astronomical Society, 492, 1855, \dodoi{10.1093/mnras/stz3572}

\bibitem[{Most \& Wang(2024)}]{most_magnetically_2024}
Most, E.~R., \& Wang, H.-Y. 2024, The Astrophysical Journal Letters, 973, L19, \dodoi{10.3847/2041-8213/ad7713}

\bibitem[{Naethe~Motta {et~al.}(2025)Naethe~Motta, Jacquemin-Ide, Nemmen, Liska, \& Tchekhovskoy}]{naethe_motta_black_2025}
Naethe~Motta, P., Jacquemin-Ide, J., Nemmen, R., Liska, M. T.~P., \& Tchekhovskoy, A. 2025, Black hole spectral states revealed in {GRMHD} simulations with texture memory accelerated cooling,  arXiv, \dodoi{10.48550/arXiv.2505.08855}

\bibitem[{Noble {et~al.}(2006)Noble, Gammie, McKinney, \& Del~Zanna}]{noble_primitive_2006}
Noble, S.~C., Gammie, C.~F., McKinney, J.~C., \& Del~Zanna, L. 2006, The Astrophysical Journal, 641, 626, \dodoi{10.1086/500349}

\bibitem[{Noble {et~al.}(2009)Noble, Krolik, \& Hawley}]{noble_direct_2009}
Noble, S.~C., Krolik, J.~H., \& Hawley, J.~F. 2009, The Astrophysical Journal, 692, 411, \dodoi{10.1088/0004-637X/692/1/411}

\bibitem[{Ogilvie \& Pringle(1996)}]{ogilvie_non-axisymmetric_1996}
Ogilvie, G.~I., \& Pringle, J.~E. 1996, Monthly Notices of the Royal Astronomical Society, 279, 152, \dodoi{10.1093/mnras/279.1.152}

\bibitem[{Padovani(2011)}]{padovani_microjansky_2011}
Padovani, P. 2011, Monthly Notices of the Royal Astronomical Society, 411, 1547, \dodoi{10.1111/j.1365-2966.2010.17789.x}

\bibitem[{Parfrey {et~al.}(2015)Parfrey, Giannios, \& Beloborodov}]{parfrey_black_2015}
Parfrey, K., Giannios, D., \& Beloborodov, A.~M. 2015, Monthly Notices of the Royal Astronomical Society: Letters, 446, L61, \dodoi{10.1093/mnrasl/slu162}

\bibitem[{Porth {et~al.}(2021)Porth, Mizuno, Younsi, \& Fromm}]{porth_flares_2021}
Porth, O., Mizuno, Y., Younsi, Z., \& Fromm, C.~M. 2021, Monthly Notices of the Royal Astronomical Society, 502, 2023, \dodoi{10.1093/mnras/stab163}

\bibitem[{Ressler {et~al.}(2015)Ressler, Tchekhovskoy, Quataert, Chandra, \& Gammie}]{ressler_electron_2015}
Ressler, S.~M., Tchekhovskoy, A., Quataert, E., Chandra, M., \& Gammie, C.~F. 2015, Monthly Notices of the Royal Astronomical Society, 454, 1848, \dodoi{10.1093/mnras/stv2084}

\bibitem[{Ressler {et~al.}(2017)Ressler, Tchekhovskoy, Quataert, \& Gammie}]{ressler_disc-jet_2017}
Ressler, S.~M., Tchekhovskoy, A., Quataert, E., \& Gammie, C.~F. 2017, Monthly Notices of the Royal Astronomical Society, 467, 3604, \dodoi{10.1093/mnras/stx364}

\bibitem[{Ripperda {et~al.}(2022)Ripperda, Liska, Chatterjee, Musoke, Philippov, Markoff, Tchekhovskoy, \& Younsi}]{ripperda_black_2022}
Ripperda, B., Liska, M., Chatterjee, K., {et~al.} 2022, The Astrophysical Journal, 924, L32, \dodoi{10.3847/2041-8213/ac46a1}

\bibitem[{Rothstein \& Lovelace(2008)}]{rothstein_advection_2008}
Rothstein, D.~M., \& Lovelace, R. V.~E. 2008, The Astrophysical Journal, 677, 1221, \dodoi{10.1086/529128}

\bibitem[{Scepi {et~al.}(2021)Scepi, Begelman, \& Dexter}]{scepi_magnetic_2021}
Scepi, N., Begelman, M.~C., \& Dexter, J. 2021, Monthly Notices of the Royal Astronomical Society, 502, L50, \dodoi{10.1093/mnrasl/slab002}

\bibitem[{Scepi {et~al.}(2023)Scepi, Begelman, \& Dexter}]{scepi_magnetic_2023}
---. 2023, Magnetic support, wind-driven accretion, coronal heating, and fast outflows in a thin magnetically arrested disc, \dodoi{10.48550/arXiv.2302.10226}

\bibitem[{Scepi {et~al.}(2022)Scepi, Dexter, \& Begelman}]{scepi_sgr_2022}
Scepi, N., Dexter, J., \& Begelman, M.~C. 2022, Monthly Notices of the Royal Astronomical Society, 511, 3536, \dodoi{10.1093/mnras/stac337}

\bibitem[{Scepi {et~al.}(2024)Scepi, Dexter, Begelman, Marcel, Ferreira, \& Petrucci}]{scepi_thermal_2024}
Scepi, N., Dexter, J., Begelman, M.~C., {et~al.} 2024, Astronomy and Astrophysics, 692, A153, \dodoi{10.1051/0004-6361/202451568}

\bibitem[{Scepi {et~al.}(2020)Scepi, Lesur, Dubus, \& Jacquemin-Ide}]{scepi_magnetic_2020}
Scepi, N., Lesur, G., Dubus, G., \& Jacquemin-Ide, J. 2020, Astronomy and Astrophysics, 641, A133, \dodoi{10.1051/0004-6361/202037903}

\bibitem[{Shakura \& Sunyaev(1973)}]{shakura_black_1973}
Shakura, N.~I., \& Sunyaev, R.~A. 1973, Astronomy and Astrophysics, 24, 337.
\newblock \url{http://cdsads.u-strasbg.fr/abs/1973A%26A....24..337S}

\bibitem[{Spruit {et~al.}(1995)Spruit, Stehle, \& Papaloizou}]{spruit_interchange_1995}
Spruit, H.~C., Stehle, R., \& Papaloizou, J. C.~B. 1995, Monthly Notices of the Royal Astronomical Society, 275, 1223, \dodoi{10.1093/mnras/275.4.1223}

\bibitem[{Sun {et~al.}(2018)Sun, Yang, Rieger, Liu, \& Aharonian}]{sun_energy_2018}
Sun, X.-N., Yang, R.-Z., Rieger, F.~M., Liu, R.-Y., \& Aharonian, F. 2018, Astronomy and Astrophysics, 612, A106, \dodoi{10.1051/0004-6361/201731716}

\bibitem[{Tchekhovskoy {et~al.}(2011)Tchekhovskoy, Narayan, \& McKinney}]{tchekhovskoy_efficient_2011}
Tchekhovskoy, A., Narayan, R., \& McKinney, J.~C. 2011, Monthly Notices of the Royal Astronomical Society: Letters, 418, L79, \dodoi{10.1111/j.1745-3933.2011.01147.x}

\bibitem[{White {et~al.}(2019)White, Stone, \& Quataert}]{white_resolution_2019}
White, C.~J., Stone, J.~M., \& Quataert, E. 2019, The Astrophysical Journal, 874, 168, \dodoi{10.3847/1538-4357/ab0c0c}

\bibitem[{Zamaninasab {et~al.}(2014)Zamaninasab, Clausen-Brown, Savolainen, \& Tchekhovskoy}]{zamaninasab_dynamically_2014}
Zamaninasab, M., Clausen-Brown, E., Savolainen, T., \& Tchekhovskoy, A. 2014, Nature, 510, 126, \dodoi{10.1038/nature13399}

\bibitem[{Zhang {et~al.}(2025)Zhang, Stone, White, Davis, Jiang, \& Mullen}]{zhang_radiation_2025}
Zhang, L., Stone, J.~M., White, C.~J., {et~al.} 2025, eprint arXiv:2509.10638, arXiv:2509.10638, \dodoi{10.48550/arXiv.2509.10638}

\bibitem[{Zhdankin {et~al.}(2023)Zhdankin, Ripperda, \& Philippov}]{zhdankin_particle_2023}
Zhdankin, V., Ripperda, B., \& Philippov, A.~A. 2023, Physical Review Research, 5, 043023, \dodoi{10.1103/PhysRevResearch.5.043023}

\bibitem[{Zhu \& Stone(2018)}]{zhu_global_2018}
Zhu, Z., \& Stone, J.~M. 2018, The Astrophysical Journal, 857, 34, \dodoi{10.3847/1538-4357/aaafc9}

\bibitem[{Zimniak {et~al.}(2024)Zimniak, Ferreira, \& Jacquemin-Ide}]{zimniak_influence_2024}
Zimniak, N., Ferreira, J., \& Jacquemin-Ide, J. 2024, Astronomy and Astrophysics, 692, A99, \dodoi{10.1051/0004-6361/202450501}

\end{thebibliography}

\end{document}